\documentclass[preprint,12pt]{elsarticle}

\usepackage{amssymb}
\usepackage{amsmath}
\usepackage{multirow}
\usepackage{placeins}

\journal{Optics \& Laser Technology}

\begin{document}

\begin{frontmatter}



\title{Synergizing Deep Learning and Phase Change Materials for Four-state Broadband Multifunctional Metasurfaces in the Visible Range } 


\author[label1,label3,label4]{Md. Ehsanul Karim} 
\author[label2,label4]{Md. Redwanul Karim} 
\author{Sajid Muhaimin Choudhury\fnref{label1}}
\ead{sajid@eee.buet.ac.bd}

\affiliation[label1]{organization={Department of Electrical and Electronic Engineering, Bangladesh University of Engineering and Technology},
            city={Dhaka},
            postcode={1205}, 
            country={Bangladesh}}

\affiliation[label2]{organization={Department of Computer Science and Engineering, Bangladesh University of Engineering and
Technology},
            city={Dhaka},
            postcode={1205}, 
            country={Bangladesh}}

\affiliation[label3]{organization={Department of Electrical and Electronic Engineering, BRAC University},
            city={Dhaka},
            postcode={1212}, 
            country={Bangladesh}}

\affiliation[label4]{Co-first authors with equal contributions}

\begin{abstract}
In this article, we report, for the first time, broadband multifunctional metasurfaces with more than four distinct functionalities. The constituent meta-atoms combine two different phase change materials, $\mathrm{VO_2}$ and $\mathrm{Sb_2S_3}$ in a multi-stage configuration. Finite-difference time-domain (FDTD) simulations demonstrate a broadband reflection amplitude switching between the four states in visible range due to the enhanced cavity length modulation effect from the cascaded Fabry-Perot cavities, overcoming the inherent small optical contrast between the phase change material (PCM) states. This, along with the reflection phase control between the four states, allows us to incorporate both amplitude and phase-dependent properties in the same metasurface - achromatic deflection, wavelength beam splitting, achromatic focusing, and broadband absorption, overcoming the limitations of previous functionality switching mechanisms for the visible band. We have used a Tandem Neural network-based inverse design scheme to ensure the stringent requirements of different states are realized. We have used two forward networks for predicting the reflection amplitude and phase for a meta-atom within the pre-defined design space. The excellent prediction capability of these surrogate models is utilized to train the reverse network. The inverse design network, trained with a labeled data set, is capable of efficiently navigating through the vast parametric space to produce the optimized meta-units given the desired figure-of-merits in terms of reflection amplitude and phase for the four states. The optical characteristics of two inverse-designed metasurfaces have been evaluated as test cases for two different sets of design parameters in the four states. Both structures demonstrate the four desired broadband functionalities while closely matching the design requirements, suggesting their potential in visible-range portable medical imaging devices. The findings of this work will open new horizons for active metasurface design with broadband complex multifunctionalities in different spectral bands.  
\end{abstract}




\end{frontmatter}


\section{Introduction}
Metasurfaces because of their distinct ability to simultaneously mold the phase, amplitude, and polarization of electromagnetic (EM) waves \cite{kim2022tunable}, have been integrated into several nanophotonic applications, namely focusing \cite{zhuang2019high}, cloaking \cite{chu2018hybrid}, absorption \cite{guo2016ultra}, holography \cite{wan2017metasurface}, beam steering \cite{wei2013highly}. Recently, there has been a growing interest in integrating several of these functionalities into the same meta-device to keep pace with the ever-increasing demand for miniaturization and integration in complex photonic systems. There have been reports of several multifunctional metasurfaces in different spectral bands utilizing graphene, phase change materials (PCM),  micro-electromechanical system (MEMS), varactor, and PIN diodes to switch between different functional states \cite{gao2020programmable,xu2022multifunctional,wang2023multifunctional}. These works have, however, mostly been limited to the near and mid-IR range owing to several limitations of the switching mechanisms. The popularly used PIN diode approach for individual control of unit cells in previous multifunctional metasurfaces is unsuitable for the visible spectrum devices with nano-scale meta-units \cite{lin2022coding}. The Graphene and MEMS-based approach relies on the precise control of Fermi level and mechanical force, respectively, and are mainly limited to the IR regime \cite{badloe2021tunable}.  

PCMs like $\mathrm{Ge_2Sb_2Te_5}$ (GST), $\mathrm{Ge_2Sb_2Se_4Te_1}$(GSST), Vanadium dioxide ($\mathrm{VO_2}$), Antimony Sulfide ($\mathrm{Sb_2S_3}$), due to their inter-state optical contrast, fast repeatable switching, and chemical stability have been used to produce broadband reconfigurable multifunctionality mainly in the IR and microwave portion of the EM spectra \cite{ma2021phase,chen2019switchable,zhang2022multifunctional,zhou2024optimized,bai2019near}. However, the transition of such broadband switching behavior towards visible frequencies is difficult due to the relatively small optical contrast between the PCM states in this band. Recently, we proposed a broadband bi-functional metasurface in the visible band harnessing cavity length modulation effect from $\mathrm{VO_2}$ phase transition \cite{karim2023reconfigurable}. However, metasurfaces capable of producing three or more different functionalities (either at a single or broad wavelength regime) in the visible range remain unrealized to date. This is due to the fact that PCMs possess only two stable states, and other functionality-switching mechanisms are ineffective or impractical in the visible range. This issue may be alleviated by combining different PCMs in the same metasurface to produce more than two states. However, previous works involving such meta-devices have mainly been limited to peak wavelength tuning \cite{chen2023multistate,meng2021four,xie2024multistage} or single-wavelength amplitude switching \cite{meng2018design} between different states.   

Broadband (or achromatic) functionalities are, in general, difficult to design in any spectral band. The commonly used forward design approach involves choosing meta-units capable of simultaneously satisfying position and frequency-dependent phase and amplitude requirements through time-consuming simulation over large parametric space is impractical for such device design. This is one of the major reasons why all the previous works showcasing four or more distinct functions in the same broadband meta-device have at least one function that works only for a single wavelength and hence is not broadband in the true sense \cite{li2019design,zhu2021four,yang2022switchable,xu2022multifunctional,ma2022highly,wang2024design,zhang2023switchable}. To the best of our knowledge, a truly broadband metasurface combining four or more functionalities has not been reported in any spectral band. 

Different numerical optimization techniques have been widely incorporated in nanophotonics to avoid the bottlenecks of the so-called brute-force design approach \cite{elsawy2020numerical}. Traditional optimization techniques (Particle Swarm Optimization, Genetic algorithm, etc.) are incapable of generalizing the complex relationship between nanostructure topologies and their spectral response \cite{gao2022inverse}. Such optimizers are reset each time for a new target, with no prior knowledge of previous cycles, making such iterative inverse design methods inefficient for multifunctional device design \cite{elesin2014time}. Deep learning-based algorithms can generate ultra-fast solutions for a given target response once the computationally expensive training using the sample dataset is over. There have been several reports of utilizing neural network architectures for inverse designing multifunctional metasurfaces \cite{an2021multifunctional,kiani2022conditional,ma2022pushing,zhu2020multifunctional}. However, they have mostly been limited to Generative Neural Networks, namely variant autoencoder (VAE) and generative adversarial network (GAN). VAE and GAN architectures almost invariably require large training sample data and are incapable of mapping 3d geometries \cite{noureen2022unique}. Another architecture that has gained popularity among the photonics research community for the inverse design of meta-devices is the Tandem Neural Network (TNN). TNN utilizes a surrogate model network (or spectral response predicting network) connected in tandem with an inverse design network to overcome the multiple solution issue in photonics inverse design and to ensure the generation of practically feasible structures \cite{xu2021improved}. Also, such tandem architectures have shown excellent performance in large parametric spaces with a reasonably sized data set \cite{yeung2021designing}. However, the vast potential of the TNN architectures remains unexplored for multifunctional metasurface design to date.  

In this work, we propose a novel methodology for combining four distinct broadband functionalities in the visible range, overcoming the limitations of previous functionality switching approaches. The basic building blocks of the proposed metasurfaces combine two different PCMs ($\mathrm{Sb_2S_3}$ and $\mathrm{VO_2}$) in a multistage configuration to overcome the small optical contrast of PCMs in the visible band. Such a combination of volatile and non-volatile PCMs in the same structure, along with the careful engineering of the cavity length modulation effect, allowed us to produce four distinct broadband states, switchable in terms of both reflection amplitude and phase, with aligned bands in the visible range. This unique EM response of our novel unit cells shows the potential of solving the issues involving previous functionality switching methods (PIN, varactor diode, PCM, Graphene, or MEMS) in the visible regime, as discussed above. However, the added flexibility in a single unit cell results in a substantial parametric space to navigate. Using the traditional trial and error method over this large design space based on the time-consuming numerical simulations to satisfy the multiple space and wavelength-dependent phase and amplitude requirements in the four states for desired functionalities is highly impractical. We have used a TNN-based architecture, capable of predicting structural parameters within the predefined search space given the requirements for the four states, to alleviate this issue. It consists of two forward networks for predicting the reflection amplitude and phase of a given meta-unit. The excellent ability of both the networks in unscrambling the complex relationship between geometrical topology and EM response of the untrained test data means they can substitute the time-consuming numerical simulations for training the reverse design network. The reverse network is capable of efficiently navigating the parametric space to produce optimized unit cells given a target spectral response. The predicted meta-units are then stitched together to form the final metasurface capable of demonstrating achromatic deflection, wavelength beam splitting, achromatic focusing, and broadband absorption in the four different states with properly aligned bands. We have tested the accuracy of this inverse design scheme by evaluating the optical response of two multifunctional metasurfaces with varied design parameters corresponding to the four functionalities. Numerical simulations demonstrate the close matching of the varied range of functional requirements in all four states for both test cases. Our inverse-designed broadband metasurfaces may find direct applications in visible-range medical imaging technologies like visible Optical Coherence Tomography (vis-OCT) systems. The findings of this work are expected to pave the way for broadband multifunctional metasurface devices for miniaturized nanophotonic systems in different spectral bands.
\section{Four-state unit cell design}
\label{sec:unit cell}

\begin{figure}[!h]
  \centerline{\includegraphics[width=6.5in]{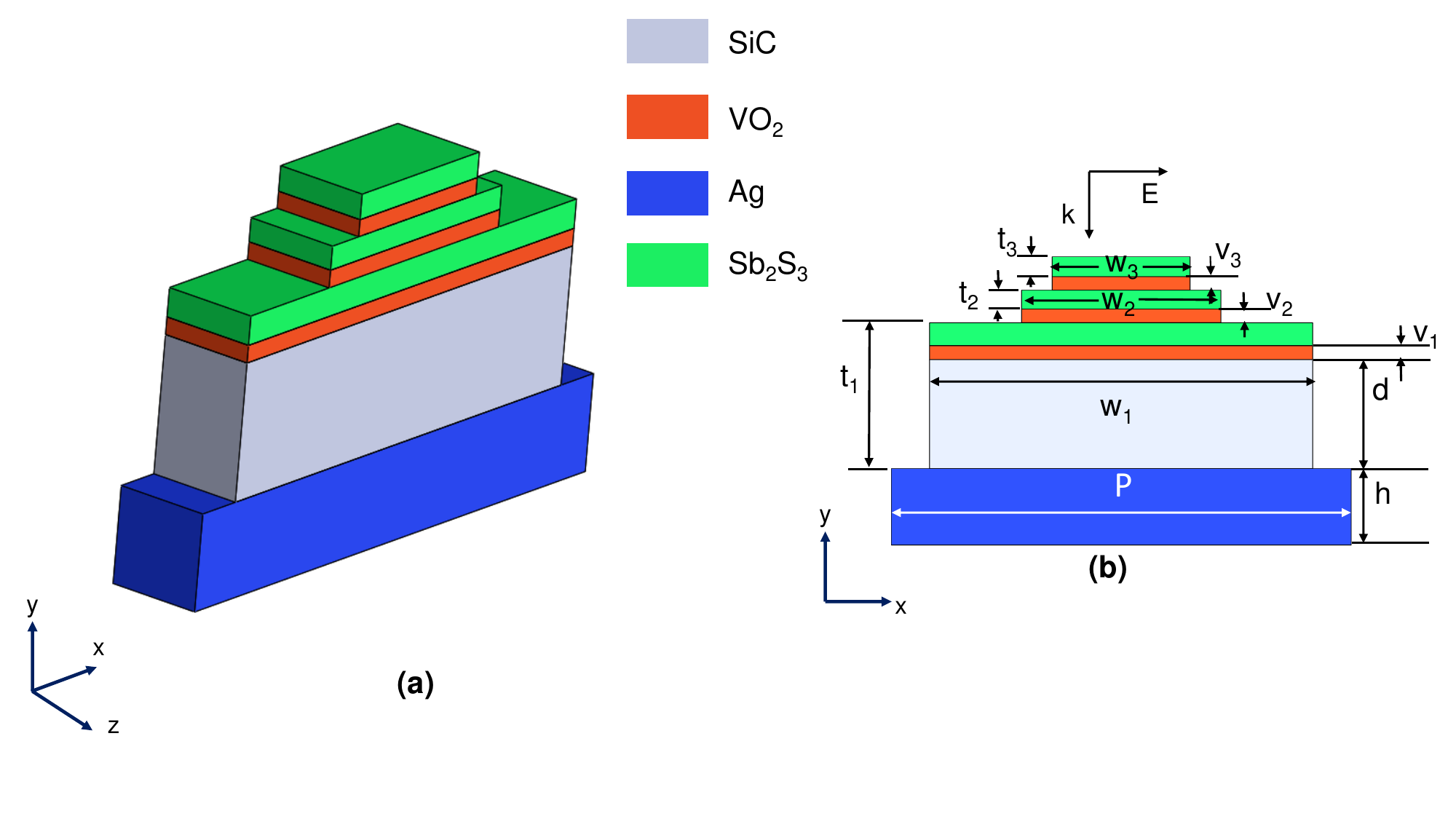}}

\caption{The \textbf{(a)} 3D schematic and \textbf{(b)} x-y plane cross-sectional illustration of our proposed 4-state unit cell with different structural parameters indicated. $P$ and $h$ represent the period of the unit cell and the height of the back-reflector (Ag), respectively. The position of the $\mathrm{VO_2}$ layer in the lowest stage relative to the Ag top surface is marked as $d$. $w_i$, $t_i$, and $v_i$ represent the cavity width, length, and $\mathrm{VO_2}$ position in the $i$th stage ($i$=1,2, and 3). The E and k-vector directions represent the polarization and propagation directions of the incident EM wave, respectively.}
\label{fig:unit_cell}
\end{figure}

Our multifunctional metasurface with four different functionalities is constructed by stitching together switchable meta-atom structures. The 3d schematic view and x-y plane cross-section of the meta-unit are illustrated in Figure \ref{fig:unit_cell}(a) and (b), respectively. The unit cell consists of three stages of cavities, each composed of $\mathrm{VO_2}$ and $\mathrm{Sb_2S_3}$ layers. The use of multiple of these stages allows us to overcome the low refractive index contrast of the PCMs in the visible regime. The phase transition of $\mathrm{VO_2}$ by temperature control between metallic and insulating phases is volatile \cite{crunteanu2015electric}. On the other hand, $\mathrm{Sb_2S_3}$ can be switched between its amorphous and crystalline states through electrical and optical stimuli \cite{delaney2020new,chen2023non}. This phase transition is non-volatile and occurs at much higher temperatures \cite{chen2023multistate} compared to the phase transition temperature of $\mathrm{VO_2}$ (68$\mathrm{^o}$C) \cite{kocer2015thermal}. Also, previous literature proves the stability of $\mathrm{VO_2}$ optical properties before and after the $\mathrm{Sb_2S_3}$ phase switching \cite{taylor2019temperature,meng2018design}. Combining volatile and non-volatile PCMs in the same structure and their stability during each other's phase transitions allow us to utilize four switchable states of the meta-atoms. This alleviates the necessity of selective control over meta-atoms for multifunctionality, a method widely used for the IR regime but difficult to integrate into devices operating at visible frequency \cite{lin2022coding,badloe2021tunable}. Several previous works show the feasibility of our proposed structure \cite{liao2018effects,pham2020versatile,cheng2019enhanced,abu2008surface,meng2018design}. However, the practical implementation is beyond the scope of this work.

\begin{figure}[!h]
  \centerline{\includegraphics[width=6.5in]{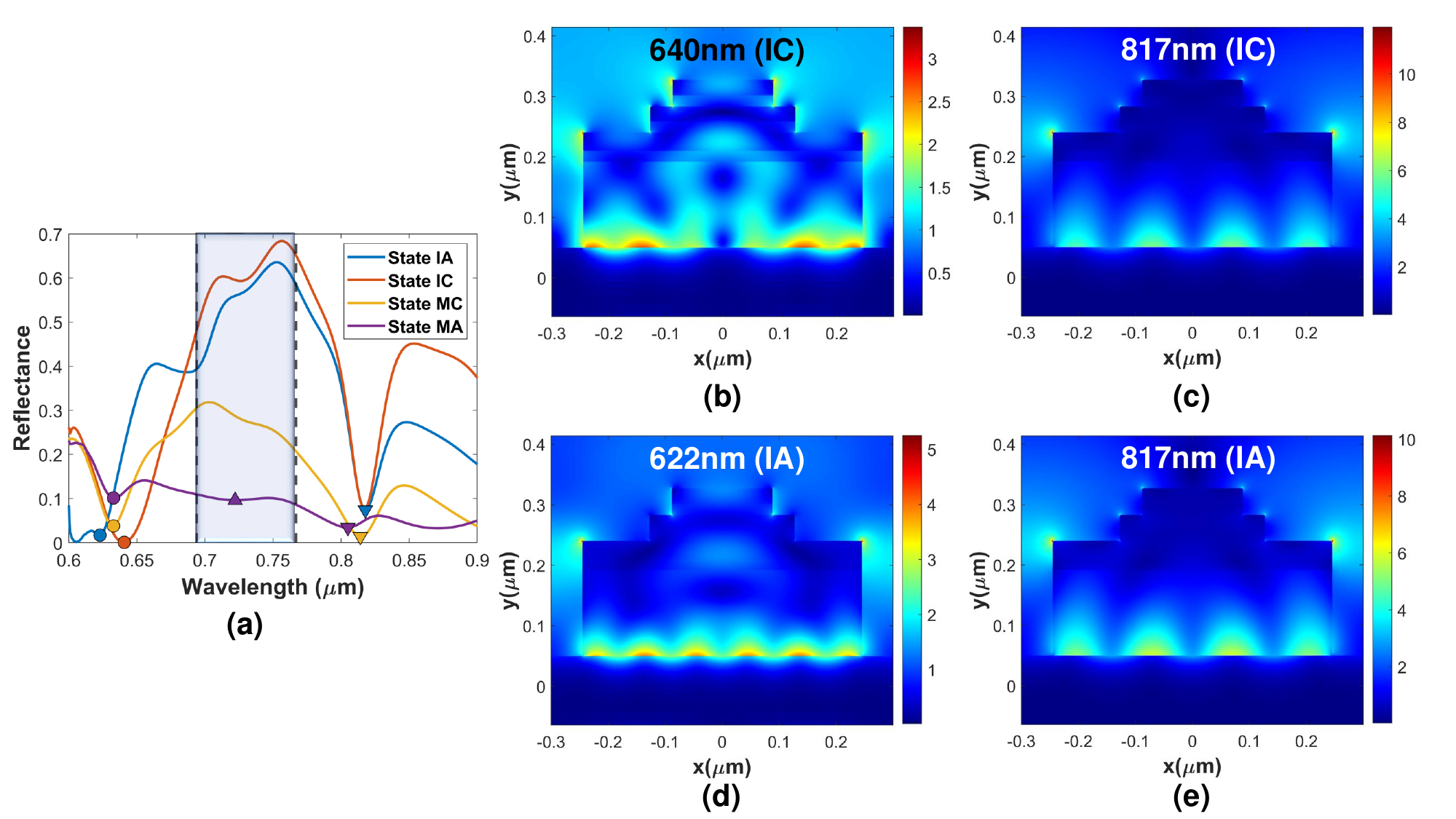}}

\caption{\textbf{(a)} The reflection spectra for four different states of the unit cell with $P$=600nm, $h$=100nm, $w_1$=490nm, $w_2$=255nm, $w_3$=176nm, $d$=142nm, $v_1$=$v_2$=$v_3$=18nm, $t_1$=190nm, $t_2$=24.6nm, and $t_3$=25.5nm. State I(or M)A(or C) represents the unit cell with the $\mathrm{VO_2}$ layers in the insulating (or metallic) state and $\mathrm{Sb_2S_3}$ layers in the amorphous (or crystalline) state. The blue-shaded region represents the working bandwidth for this unit cell configuration. The symbols $\circ$, $\bigtriangledown$, and $\bigtriangleup$ indicate the Mode 1, 2, and 3 wavelength positions, respectively, at the respective states. The x-y plane electrical field distribution of this representative unit cell for Mode 1 and 2 in the \textbf{(b-c)} IC state and \textbf{(d-e)} IA state. The total electric field magnitudes in each case have been normalized by the field magnitude of the incident wave. }
\label{fig:field1}
\end{figure}

The reflection spectra of a representative unit cell obtained from finite-difference time-domain (FDTD) simulations for all four states are shown in Figure \ref{fig:field1}(a). Details of the simulation setup can be found in the Methods Section of the Supporting Information (SI). The numerical simulation results clearly show four distinct broadband states, with effective amplitude switching between the on (MC, IC, IA) and off (MA) states. Here, State I(or M)A(or C) represents the unit cell with the $\mathrm{VO_2}$ layers in the insulating (or metallic) phase and $\mathrm{Sb_2S_3}$ layers in the amorphous (or crystalline) state. This EM response is in contrast to previous works on metasurfaces involving multiple PCMs, where either peak wavelength tuning \cite{chen2023multistate,meng2021four,xie2024multistage} or single wavelength amplitude switching \cite{meng2018design} has been demonstrated. Due to the presence of an Ag back reflector, the transmission ($T$) through the unit cell is negligible. So, we have defined the operating bandwidth in terms of the reflection amplitude or reflectance ($R$) values. We require the final metasurface to produce achromatic beam deflection, wavelength beam splitting, achromatic focusing, and broadband absorption in the IA, IC, MC, and MA states of the unit cells, respectively. The blue-shaded region in Figure \ref{fig:field1}(a) from 694nm-766.96nm represents the switchable (in terms of amplitude) operating bandwidth for this representative meta-unit. This region has a minimum of 40\% reflectance ($R$) for the IA and IC states, 20\% for the MC state, and a maximum of 10\% reflectance for the MA state (off state). These threshold values have been selected based on previous relevant literature considering the desired functionality in the respective state \cite{li2023active}. 

\begin{figure}[!h]
  \centerline{\includegraphics[width=6.5in]{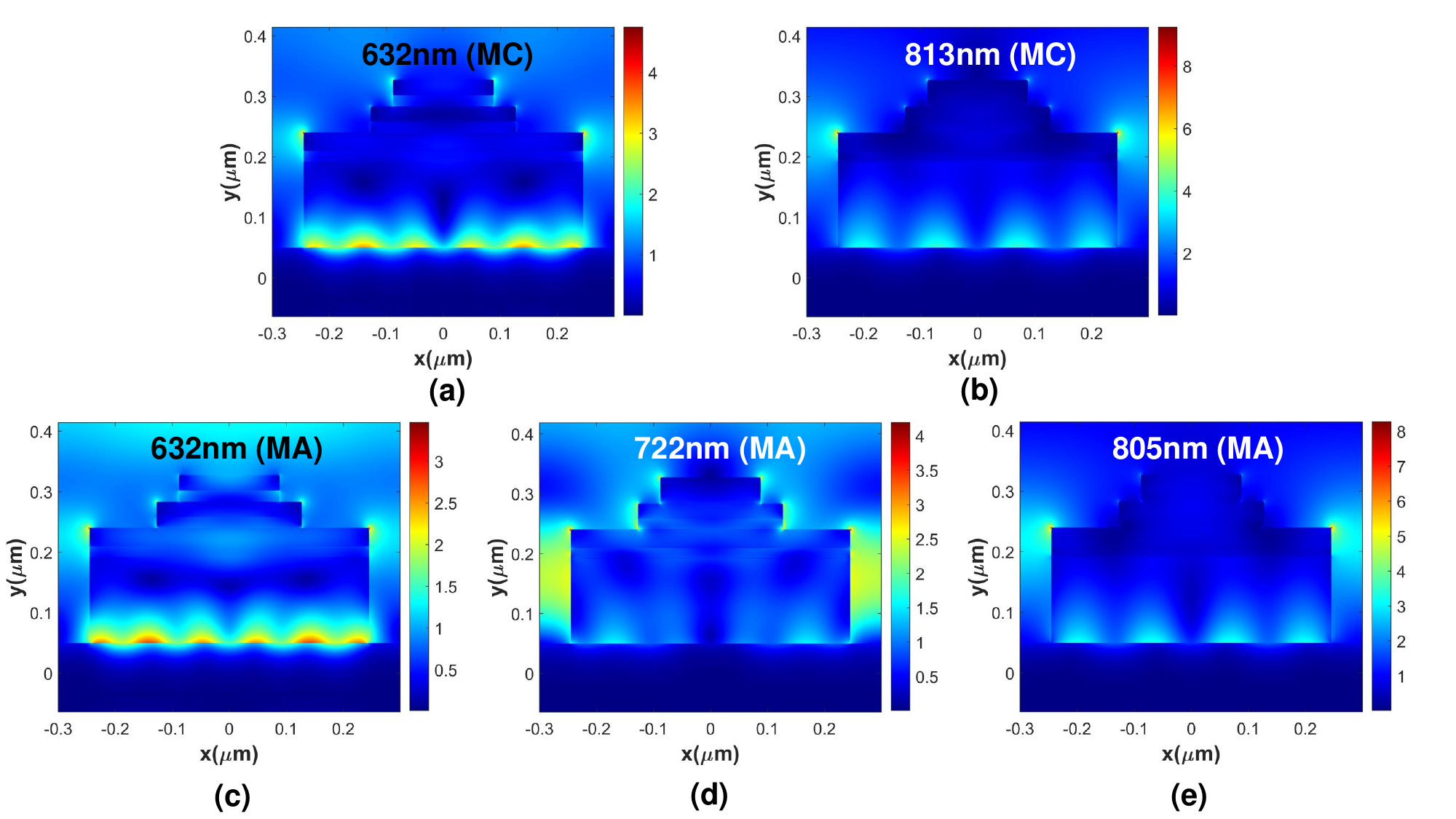}}

\caption{The x-y plane electrical field distribution of the representative unit cell (in Figure \ref{fig:field1}) for \textbf{(a-b)} Mode 1 and 2 in the MC state and \textbf{(c-e)} Mode 1, 2, and 3 in the MA state. The total electric field magnitudes in each case have been normalized by the field magnitude of the incident wave.}
\label{fig:field2}
\end{figure}

The reflection spectra of the unit cell in Figure \ref{fig:field1}(a) reveal the presence of two common minima (or resonant modes) at each of the four states. These minimum points are designated as Mode 1 and 2 as marked in Figure \ref{fig:field1}(a). An additional minima becomes prominent at 722nm for the MA state, referred to as Mode 3 in our work. This additional absorption peak results in the broadband absorption characteristics of the MA state. To get a better insight into the origin of these reflection minima, we have also simulated the x-y plane spatial electrical field profiles of the representative unit cell at all the states at their respective resonance frequencies. The results have been shown in Figure \ref{fig:field1} and \ref{fig:field2}. Field distribution for Mode 1 in both the IC and IA states (Figure \ref{fig:field1}(b), (d)) show strong field concentration in the cavity formed between the Ag back reflector and the $\mathrm{Sb_2S_3}$-air interface. This indicates that the resonances are dominated by the Fabry-Perot (FP) cavity mode. For the IA state, the field profiles indicate a continuous FP cavity between Ag and the upper $\mathrm{Sb_2S_3}$, due to the similar refractive indices of SiC, amorphous $\mathrm{Sb_2S_3}$, and insulating $\mathrm{VO_2}$ \cite{kalantari2021active}. However, for Mode 1 in the IC state, there is significant field crowding around the upper stages, indicating the formation of cavity modes between the $\mathrm{VO_2}$ and $\mathrm{Sb_2S_3}$-air interfaces \cite{karim2023reconfigurable}. For Mode 2 in IC and IA states, the field distributions illustrated in Fig. \ref{fig:field1}(c) and (e) indicate the dominance of the FP cavity in the lowest stage of the meta-device. This is further validated by the equal resonance wavelengths for the two states in Mode 2 \cite{shabani2021design}. 

The difference in the optical properties of metallic $\mathrm{VO_2}$ from SiC means a continuous cavity formation is not possible for either of MA or MC state. The similarity of resonant wavelengths and field distributions in the two states for both Mode 1 (Figure \ref{fig:field2}(a), (c)) and 2 (Figure \ref{fig:field2}(b), (e)) indicate FP cavity formed between the Ag substrate and lower $\mathrm{VO_2}$ being the dominating factor in resonance. For Mode 3 in the MA state, there is strong coupling between the neighboring cells alongside weak FP cavity mode, as can be seen in Figure \ref{fig:field2}(d). This resonance allows the state to operate as a broadband absorber and is not supported in any of the three states. Such resonances due to coupled nano-structure have been found to disappear upon change in refractive index in previous literature \cite{weng2019broadband}, which is the reason for the absence of Mode 3 in the other three states. The use of cascaded FP cavities in the unit cell produced the effective broadband amplitude switching between the on and off states (See Figure S2 of SI). Moreover, a phase switching is expected to appear due to the change in effective index with state transition. So, this multi-stage unit cell will allow us to incorporate both the phase (deflection, focusing, beam splitting) and amplitude (absorption) dependent functionalities in the same device.

\section{Training and evaluation of the tandem neural network}

\begin{figure}[!h]
  \centerline{\includegraphics[width=6.5in]{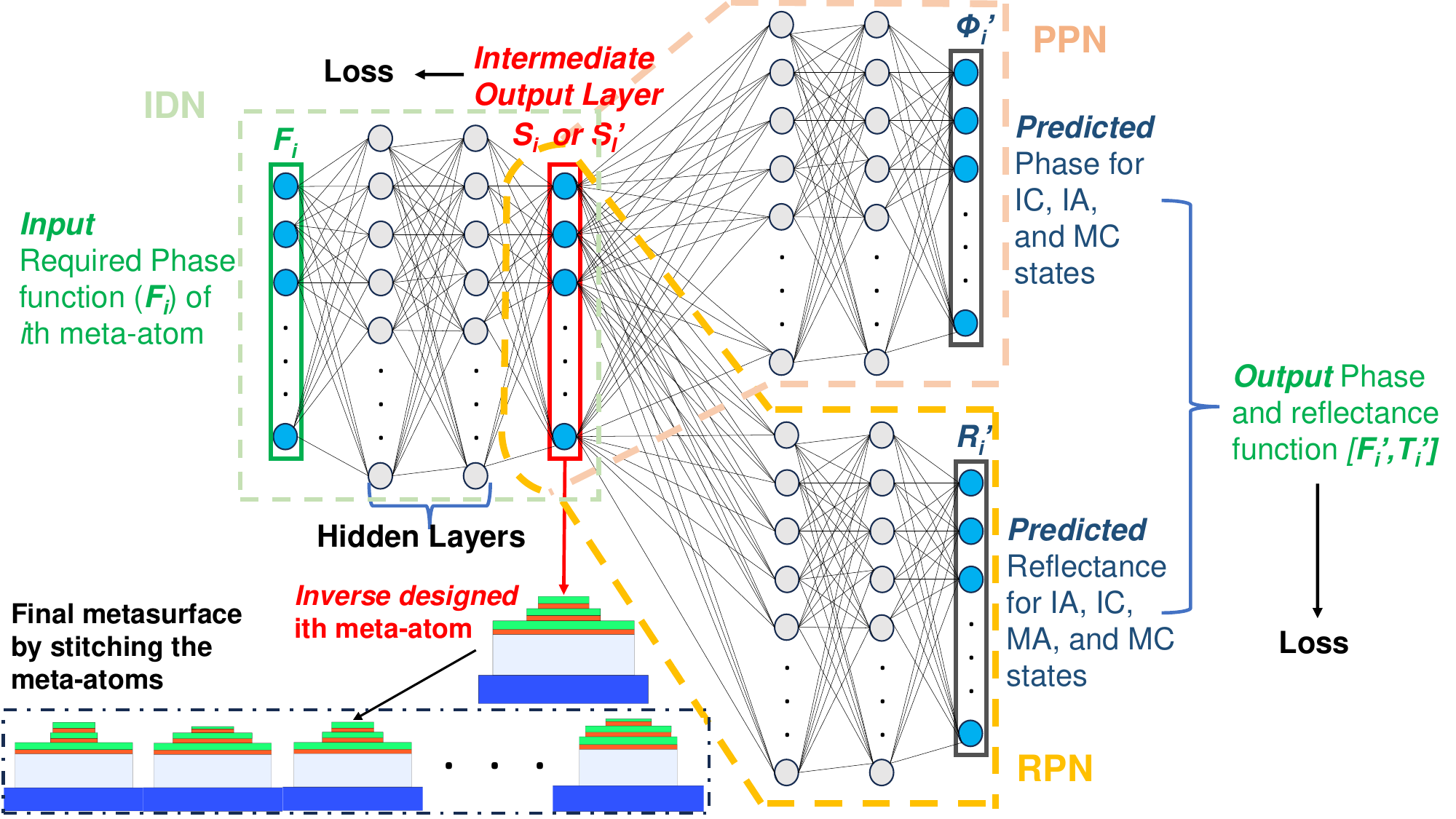}}

\caption{The inverse design methodology of the multifunctional metasurface using a Tandem Neural Network. The Inverse Design Network (IDN) optimizes the individual meta-atoms based on the phase requirements ($F$) provided as input while trying to match the reflectance (\textit{R}) thresholds for each state. The meta-atoms are stitched together to form the final multifunctional meta-device. IDN is trained using the forward networks, Phase (PPN), and Reflectance Predicting Networks (RPN), while keeping their weights fixed. The hidden layers of the three neural networks have been drawn to
illustrate the inverse design procedure and do not represent the actual architecture. }
\label{fig:Summary}
\end{figure}

Our designed tandem neural network (TNN) is schematically illustrated in Figure \ref{fig:Summary}. It can be divided into two parts, namely forward predicting and inverse design networks. The two forward networks, PPN and RPN, are trained using the labeled data collected through our simulations to solve the regression problem between the meta-atom structural parameters ($S$) and reflection phase and amplitude, respectively. The two forward networks connected in tandem with the inverse network (IDN) allow its training to produce inverse-designed meta-atoms given a set of phase requirements (\textit{F}) and reflectance thresholds for different states without encountering the multi-solution problem of nanophotonics \cite{yeung2021designing}. The final optimized IDN can inverse design individual unit cells based on the desired functionality in different states. These meta-atoms are then stitched together to form the final multifunctional meta-device, as schematically shown in Figure \ref{fig:Summary}. 

\subsection{Data set formation}
\label{sec:data}
The dataset is formed by simulating a large number of unit cells with varied structural parameters ($S$) to obtain the optical response ($D$). This will form the initial data pair ($S$, $D$) for the TNN. The simulation was conducted using Lumerical FDTD Solutions. The simulation details can be found in the Methods section of SI. $S$ represents the structure vector and is given by $S=[w, h, d, t, v_1, v_2, v_3, x_1, x_2, y_1, y_2]$. Here, we have used the parameters $x_1(=w_2/w_1)$, $x_2(=w_3/w_1)$, $y_1(=t_2/(t_1-d))$, and $y_2(=t_2/(t_1-d))$ to link $t_2$, $t_3$, $w_2$, $w_3$ with $t_1$ and $w_1$. This makes selecting the constraints on the structural parameters easier. The ranges of different parameters are $w_1$=400-550nm, $h$=50-250nm, $d$=100-280nm and $d+v_1<t_1$, $t_1$=120-300nm, $v_1$=10-30nm, $v_2$=10-30nm, $v_3$=10-30nm, $x_1$=0.1-0.9, $x_2$=0.05-0.85 and $x_2<x_1$, $y_1$=0.2-1.5, and $y_2$=0.2-1.5. These constraints have been chosen based on our experience during the unit cell design process and to avoid forming physically unrealizable structures. The period ($P$) was excluded from the independent variable list, as this would require complex stitching of meta-atoms into a single metasurface. Based on these ranges, we have randomly generated several values of the $S$ vector with enough variance. These structures are then simulated to obtain the optical characteristics ($D$) over a wavelength range of 620nm to 820nm. The frequency range is divided into 40 equally spaced points. Again, the spectral range and resolution have been chosen based on preliminary random simulations to make sure the important features of the reflection phase and amplitude can be captured adequately. For a particular structure $S_i$, the optical response $D_i=[R_{i,IA}, R_{i,IC}, R_{i,MC}, R_{i,MA}, \phi_{i,IA}, \phi_{i,IC}, \phi_{i,MC}, \phi_{i,MA}]$, a 320x1 vector. $R_{i,IA}$ and $\phi_{i,IA}$ represent the reflection amplitude and phase vectors (each a 40x1 vector) for IA state of the $i$th data point (with $S_i$ structure vector). To make the group delay calculation convenient, we have used the raw phase data from the FDTD simulations, avoiding the discrete jumps of phase between -$\pi$ and $\pi$ \cite{fan2021time}. We have collected 46,560 data points ($S_i$,$D_i$), each structured in the above-mentioned format. 

\subsection{Forward Predicting Networks}

The forward predicting networks in tandem neural networks (TNN) act as surrogate models to map the geometrical topologies of meta-atoms into corresponding optical responses. Due to the one-to-one relationship between the input and output, the forward networks are trained before the TNN architecture. We have used two separate neural networks for predicting reflection phase ($\phi$) and amplitude ($R$), namely phase predicting network (PPN) and reflectance predicting network (RPN) \cite{neto2022deep}. The supervised learning process of both networks is conducted by randomly splitting the 46,560 data points into training, validation, and test data sets with a ratio of 0.9, 0.05, and 0.05, respectively. For the RPN, the ground truth data points are ($S_i$, $R_i$), with $R_i=[R_{i,IA}, R_{i,IC}, R_{i,MC}, R_{i,MA}]$, a 160x1 vector. For the PPN, we have truncated the phase spectra of all the data points for each state to 15 points between the wavelengths 693.758nm and 769.108nm to avoid the sharp transitions in the raw phase data \cite{qiu2024neural}. The range was chosen based on our observations during the data collection and PPN training. Also, since most achromatic phase gradient metasurface applications require linear phase spectra, such truncation does not harm the total usable bandwidth. So, the data pair for PPN training is ($S_i$, $\phi_i$), with $\phi_i=[\phi_{i,IA}, \phi_{i,IC}, \phi_{i,MC}]$, a 45x1 vector. We have not included the phase response of the MA state of the unit cell to reduce the computational complexity of the final network, as this state is expected to produce broadband absorption in the final metasurface, and the phase response becomes unimportant.   

\begin{figure}[!h]
  \centerline{\includegraphics[width=6.5in]{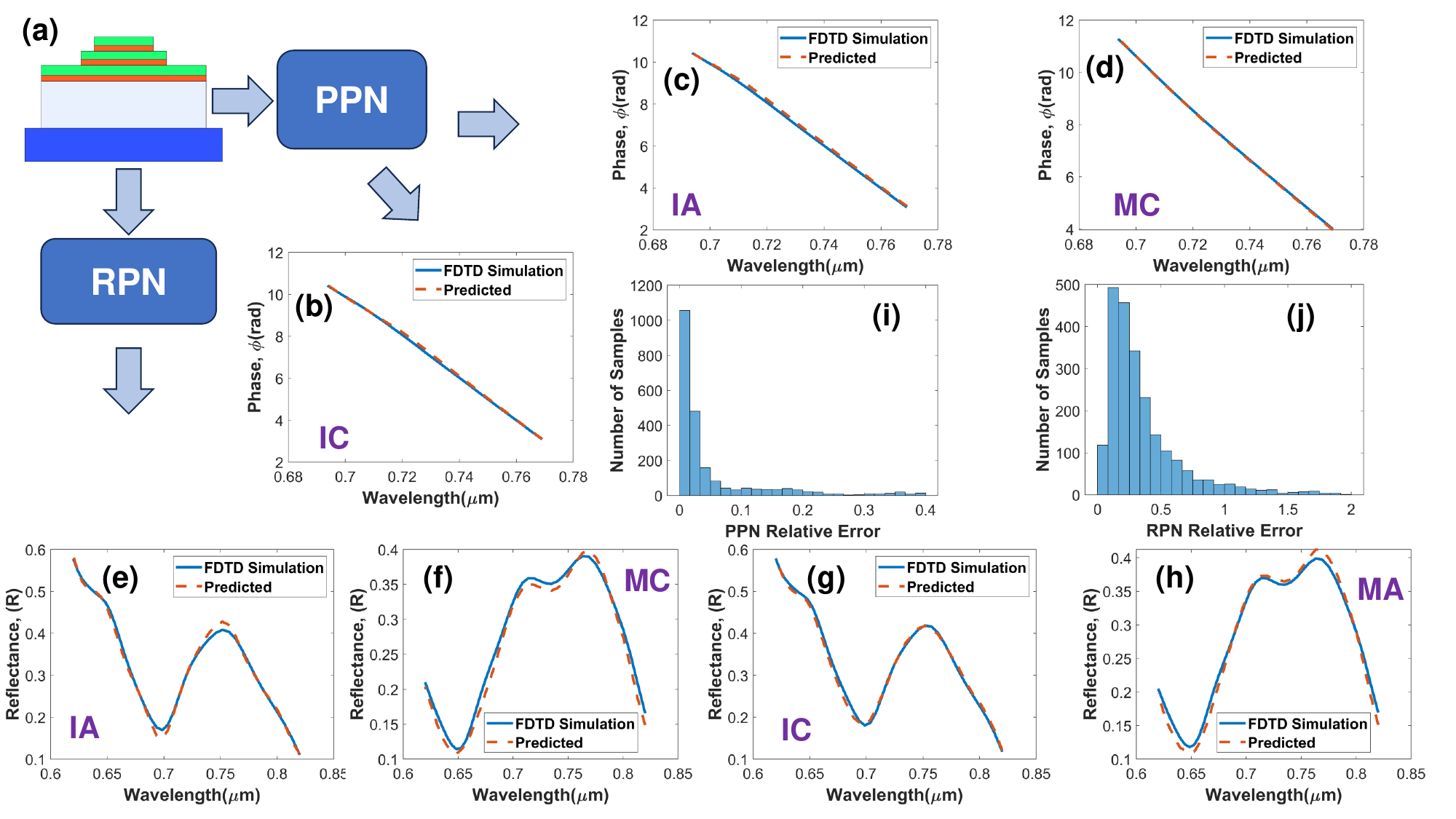}}

\caption{\textbf{(a)} Input of the two forward predicting networks, PPN and RPN, comprising 11 structural parameters. \textbf{(b)-(d)} Reflection phase ($\phi$) and \textbf{(e)-(h)} Reflectance ($R$) spectra of different states of a randomly selected unit cell from the test data set, calculated from FDTD simulations and predicted from PPN and RPN respectively. The structural parameters input to the neural networks for this test case are $w_1$=528.7849nm, $h$=111.91554nm, $d$=161.58876nm, $t_1$=187.06679nm, $v_1$=24.629309nm, $v_2$=19.88659nm, $v_3$=21.681828nm, $x_1$=0.84706473, $x_2$=0.48688328, $y_1$=1.0422835, and $y_3$=0.8203927. Relative error distributions of the \textbf{(i)} PPN and \textbf{(j)} RPN for the 2328 test samples.}
\label{fig:forward}
\end{figure}

Both the forward networks are trained on the respective training data-points. Different hyperparameters of the two networks have been chosen through the iterative training and testing of the networks so that the respective loss functions converge to a minimum without exhibiting any divergence or overfitting. The loss function for both the netowrks is Mean Absolute Error (MAE) defined as \cite{wu2021comparison} 
\begin{equation}
\label{eqn:MAE_RPN}
    \mathrm{Loss_{RPN}}=\frac{1}{N}\sum_{i=1}^{N}|R_i-R_i'|
\end{equation}
\begin{equation}
\label{eqn:MAE_PPN}
    \mathrm{Loss_{PPN}}=\frac{1}{N}\sum_{i=1}^{N}|\phi_i-\phi_i'|
\end{equation}
where $N$ is the total number of training samples, $R_i$ ($\phi_i$) and $R_i'$ ($\phi_i'$) are the ground truth and predicted values of reflectance (phase) spectra of the $i$th structure. The structural parameters were normalized before inputting into the PPN and RPN to eliminate the influence of different orders of magnitudes on the training process. The normalization was done using the well-known z-score formula \cite{mare2017nonstationary}. The final RPN and PPN architectures contain 7 (with 2000, 1600, 1200, 1000, 800, 600, 400 nodes in the layers) and 8 (with 2400, 2000, 1600, 1200, 1000, 800, 600, 400 nodes in the layers) fully connected hidden layers respectively. The optimum training of RPN was completed at 300 epochs where the validation loss converged to 0.022. For PPN, the validation loss converged to a constant value of 0.599rad after 300 training epochs, indicating effective training of the neural network (See SI). During the training process of both the networks, the learning rate was dynamically varied with decreasing loss. Different hyperparameters, training and validation loss curves and dynamic learning rates of the two neural networks can be found in the Supporting Information.    

We have tested the prediction accuracy of RPN and PPN using the untrained test data pairs. Figure \ref{fig:forward}(i) and (j) show the statistical distributions of relative errors (RE) for PPN and RPN prediction of the test data set, respectively. The relative errors for the two networks are defined as \cite{xu2021improved}
\begin{equation}
\label{eqn:RAE_RPN}
    \mathrm{RE_{RPN}}=\frac{1}{n}\sum_{j=1}^{n}|\frac{R_j-R_j'}{R_j}|
\end{equation}
\begin{equation}
\label{eqn:RAE_PPN}
    \mathrm{RE_{PPN}}=\frac{1}{n}\sum_{j=1}^{n}|\frac{\phi_j-\phi_j'}{\phi_j}|
\end{equation}
where $n$ is the number of frequency points in a test case. The RE distribution for PPN shows that 91.3\% of test data has an RE of less than 0.25. For RPN, about 90\% of the test cases have an RE smaller than 0.8. The slightly retarded prediction accuracy of the RPN is due to the small magnitudes of $R$s, resulting in large REs even for small deviations. We have also shown the reflectance and phase spectra of different states of a randomly selected unit cell from the untrained test data set predicted by RPN and PPN along with the response from simulation in Figure \ref{fig:forward}(b-h). A similar analysis for two more test cases can be found in Figures S4 and S5 of the Supporting Information. These test cases were selected from the reserved 2328 test data using the random number generator of Python. For all three cases, predictions from both forward networks are in excellent agreement with the ground truth (reflection response from FDTD simulations). The excellent prediction performance of the forward networks means they can be used in training the TNN instead of the time-consuming FDTD simulations.

\subsection{Inverse Design Network}
The goal of the IDN is to design meta-atoms with a given reflection phase response for IA, IC, and MA states while ensuring the reflectivity threshold over the design bandwidth in each of the four states. However, a stand-alone IDN suffers from poor convergence due to the so-called non-uniqueness problem in nanophotonics, caused by the fact that different meta-atoms may produce the same optical response \cite{yeung2021designing}. To resolve this issue, we have connected the IDN to our previously trained forward networks to form a TNN, as shown in Figure \ref{fig:Summary}. Previously formed training, validation, and test data are used for training the IDN. But instead of using the complete optical spectra ($R_i,\phi_i$) as the input to IDN (in other words, as the requirement for inverse designed meta-atom), we use specific requirements of the desired functionality in different states. This significantly reduces the complexity of the IDN architecture. The unit cell in the IA and MC states needs to match the required phase and group delay around a reference wavelength ($\lambda_r$=729.493nm) to achieve achromatic beam deflection and lensing, respectively. For the IC state (Wavelength beam deflection), the requirement is matching phase values at the two target wavelengths ($\lambda_1$=751.623nm and $\lambda_2$=708.629) (See SI for detailed equations). So, the phase requirements (or input of the IDN) for meta-atom inverse design is $F_i=[\phi_{i,IA}(\lambda_r), \phi_{i,IC}(\lambda_1), \phi_{i,IC}(\lambda_2), \phi_{i,MC}(\lambda_r), GD_{i,IA}, GD_{i,MC}]=[\psi_i, GD_{i}]$. Here, $\phi_{i,j}(\lambda_k)$ represents the reflection phase (wrapped between 0 to 2$\pi$) of the $j$th state ($j$= IA, IC, or MC) of the $i$th meta-atom in the data set at the wavelength $\lambda_k$ ($k=r,1,2$). $GD_{i,j}$ is the group delay in the $j$th state around $\lambda_r$ for a bandwidth between 698.65 to 763.19nm. ($F_i, S_i$) forms the 46,560 ground truth data pairs for IDN training.  

The hyperparameters are again chosen through manual adjustment to ensure the minimum value of loss function at convergence during iterative training and testing of the network. The loss function is defined as
\begin{equation}
\label{eqn:loss_IDN}
\begin{aligned}
\mathrm{Loss_{IDN}}={} & \frac{1}{2}(\frac{1}{N}\sum_{i=1}^{N}min(|\psi_i-\psi_i'|,|\psi_{i}-\psi_{i}'\pm 2\pi|)+ a_1\frac{1}{N}\sum_{i=1}^{N}T_i'+ \\
& a_2\frac{1}{N}\sum_{i=1}^{N}|GD_i-GD_i'| 
+ \frac{1}{N}\sum_{i=1}^{N}|S_i-S_i'|)
\end{aligned}
\end{equation}

where $S_i'$ is the inverse designed structural parameters output from the IDN. $F_i'=[\psi_{i}', GD_{i}']$ are the values calculated from the phase ($\phi_i'$) predicted by the PPN, as schematically shown in Figure \ref{fig:Summary}. Including the error in predicting the labeled structural parameters in the loss function restricts IDN from producing meta-atoms with structural parameters within the given constraints listed in previous sections \cite{xu2021improved}. Also, the loss component for reflectivity, $T_i'$, is calculated from the reflectance spectra ($R_i'$) predicted by the RPN. This quantity is defined as errors in $R_i$ in not being able to meet the given threshold for respective states within the design bandwidth (698.65 to 763.19nm). The detailed equations for calculating $T_i'$ can be found in the Methods Section of SI. The weighting factors $a_1$ and $a_2$ in Eq. \ref{eqn:loss_IDN} have been set at 0.02 and 0.1, respectively. The final IDN architecture has nine hidden layers, with 2000, 1600, 1200, 1000, 800, 600, 400, 200, and 100 neurons in each layer. 

\begin{table}
    \centering
\caption{IDN predicted and target structural parameters of three randomly chosen unit cells from the test data}
\label{table:RDN}
    \begin{tabular}{c|cc|cc|cc} \hline 
         \multirow{2}{5em}{Structural Parameters}&  \multicolumn{2}{|c|}{\textbf{Test Case-1}}&  \multicolumn{2}{|c|}{\textbf{Test Case-2}}&  \multicolumn{2}{c}{\textbf{Test Case-3}}
         \\
         
         &  Target&  Predicted&  Target&Predicted&  Target&Predicted\\ \hline 
         
         $w_1$(nm) & 528.7849&  535.29285 & 516.9677 & 513.103 & 516.24695 & 518.13855\\ 
         $h$(nm)& 111.91554 & 108.66646 & 70.32359 & 69.940414 & 92.14034 & 99.95201\\  
         $d$(nm)& 161.58876 & 160.80196 & 228.0329 & 218.74953 & 207.81909 & 196.51451\\  
         $t_1$(nm)& 187.06679 & 191.21019 & 252.03937 & 253.79633 & 238.82242 & 236.34608\\  
         $v_1$(nm)& 24.629309 & 24.144024 & 22.80398 & 16.024605 & 14.516156 & 18.211609\\  
         $v_2$(nm)& 19.88659 & 24.937346 & 21.69389 & 15.617595 & 22.883387 & 17.137606\\  
         $v_3$(nm)& 21.681828 & 19.810368 & 14.563346 & 19.093914 & 13.8248415 & 18.387709\\  
         $x_1$& 0.84706473 & 0.68027246 & 0.18387112 & 0.74375665 & 0.47292465 & 0.69418085\\ 
         $x_2$& 0.48688328 & 0.45067346 & 0.13597298 & 0.093863845 & 0.09616624 & 0.20294511\\
         $y_1$& 1.0422835 & 0.8832304 & 0.9895365 & 0.50954455 & 0.7904597 & 0.6452963\\
         $y_2$& 0.8203927 & 1.0613561 & 1.0169164 & 0.6326489  & 1.3373394  & 0.6581082 \\
         \hline
    \end{tabular}

\end{table}

The reverse network was trained on the 41940 training data pairs ($F_i$,$S_i$) while tandemly connected to the two forward networks (RPN and PPN). During the training process, only the weights of IDN neurons are adjusted, leaving the weights of trained RPN and PPN architectures constant. The final IDN architecture was optimally trained in 200 epochs, at the end of which the training and validation losses are 3.39 and 5.03, respectively. The learning rate was dynamically adjusted with decreasing loss during the training process. The training and validation loss curves, dynamic learning rate, and optimum settings of different hyperparameters for this neural network can be found in the SI. The performance of the IDN as an optimization tool was tested on the 2328 untrained test data set. The statistical histogram of RE for the IDN in predicting the correct structural parameters for the test cases is shown in Figure S3 of SI. This quantity is defined as
\begin{equation}
\label{eqn:MAE_IDN}
    \mathrm{RE_{IDN}}=\frac{1}{k}\sum_{j=1}^{k}|\frac{S_j-S_j'}{S_j}|
\end{equation}

where, \textit{k} is the total number of parameters (11) in the $S_j$ vector. About 92.3\% of the test cases show an error of less than 0.5. Table \ref{table:RDN} compares the structural parameters predicted by IDN with ground truth data for three test cases. These are the same test data points randomly chosen for the forward networks. The predictions are in very good agreement with the target values, proving the potential of the designed TNN architecture to produce inverse-designed broadband multifunctional metasurface with desired functionalities.

\section{Optical characteristics of the inverse designed metasurfaces}
\begin{figure}[!h]
  \centerline{\includegraphics[width=6.5in]{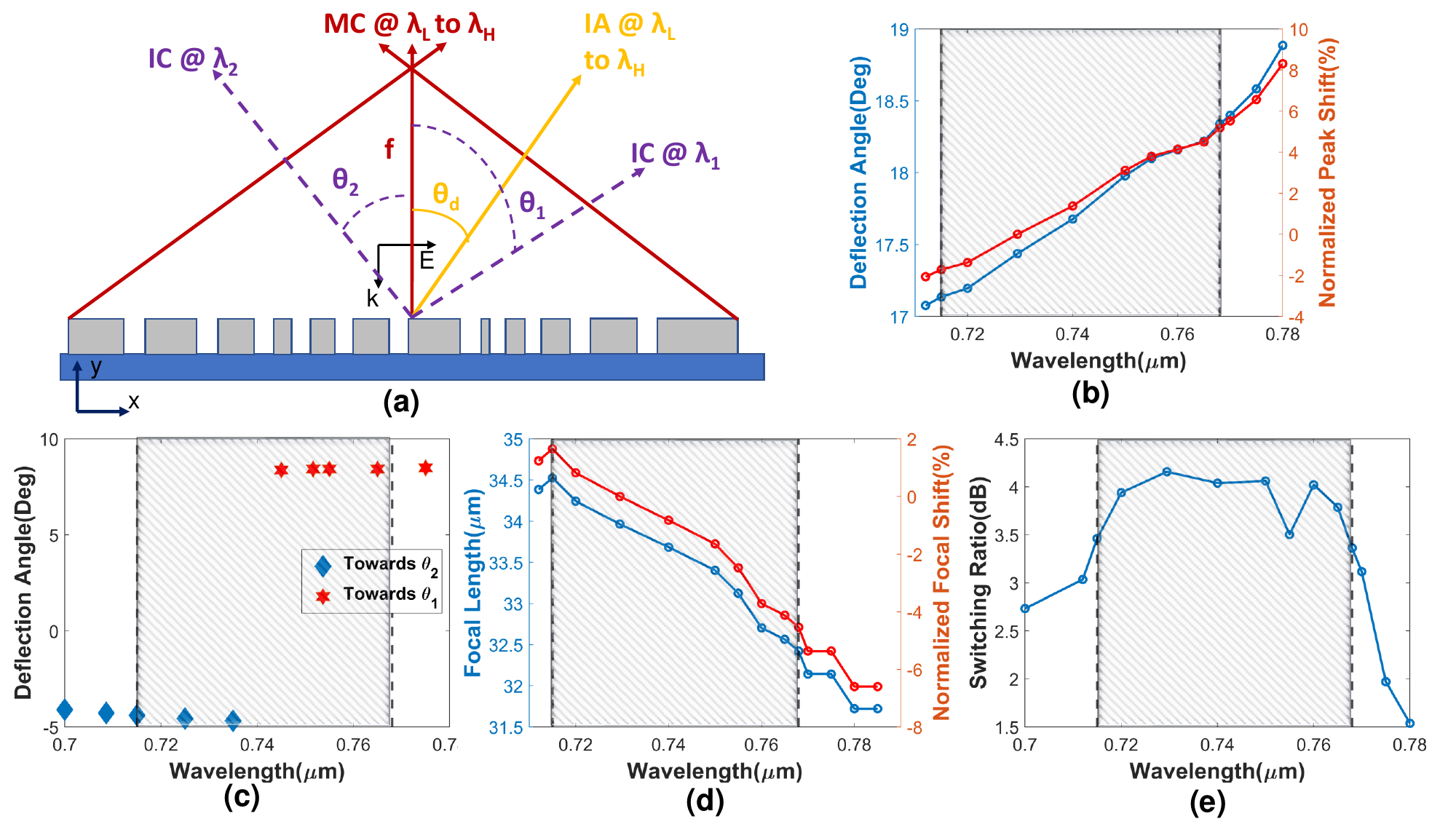}}

\caption{\textbf{(a)} Schematic representation of the operation of the inverse designed broadband multifunctional metasurfaces with an operating bandwidth ranging from $\lambda_L$ to $\lambda_H$. The meta-device achromatically deflects and focuses the reflected beam towards an angular direction of $\theta_d$ in the IA state and focal length $f$ in the MC state, respectively. In the IC state the beams with wavelength $\lambda_1$ and $\lambda_2$ get split towards angles $\theta_1$ and $\theta_2$ (with $\lambda_L\leq\lambda_1,\lambda_2\leq\lambda_H$) respectively. For the MA state, the incident wave gets absorbed by the metasurface. \textbf{(b)-(e)} Different optical parameters for the four states of BMFM-1, inverse designed for $\theta_d=17.0194^o, \theta_1=8.0237^o, \theta_2=-3.7697^o, f=32.6665\mu m, \lambda_1=751.623nm, \lambda_2=708.629nm.$ The hatched regions represent the operating wavelength between 715nm (=$\lambda_L$) and 768nm (=$\lambda_H$).}
\label{fig:4_12_summary}
\end{figure}

We have tested our TNN-based inverse design methodology by evaluating the optical performance of the optimized metasurface in the four states. We expect the predicted structure from TNN to produce achromatic deflection, wavelength beam splitting, achromatic focusing, and broadband absorption in the IA, IC, MC, and MA states, with aligned operating bands. The operating principle of such broadband multifunctional metasurface (BMFM) is illustrated schematically in Figure \ref{fig:4_12_summary}(a). The design process starts with forming the required phase function vectors (\textit{F}) for different positions on the BMFM. The calculation is based on the phase-gradient metasurface design equations for specific requirements of $\theta_d$, $\theta_1$, $\theta_2$, and $f$ (See Methods section of SI for the detailed equations). The values of $\lambda_1$, $\lambda_2$, and $\lambda_r$ have been kept fixed at 751.623nm, 708.629nm, and 729.493nm, respectively for the requirement calculations. These values are the same as in the training process. The \textit{F} vectors are input to the IDN, and the corresponding vectors for the structural parameters (\textit{S}) are obtained from the intermediate output layer of the IDN, as illustrated in Figure \ref{fig:Summary}. These inverse-designed meta-atoms are then stitched together to form the final BMFM. We then characterize the optical response of the constructed meta-device against the requirements in different states. The details of the FDTD simulations of the BMFM structure can be found in the Methods section of SI.      

The first metasurface (BMFM-1) is designed for $\theta_d=17.0194^o, \theta_1=8.0237^o, \theta_2=-3.7697^o$, and $f=32.6665\mu m$. The FDTD simulation response of the predicted (by IDN) meta-units has been compared with the required values in Figure S6 of the Supporting Information. Despite the wide range of requirements at the same point of the metasurface in different states, the response matches the requirements very closely, proving the efficiency of our TNN-based inverse design approach for BMFMs. Different optical performance parameters of the BMFM-1 are shown in Figure \ref{fig:4_12_summary}(b)-(e). In the IA state, the deflection angle ($\theta_{IA}$) for the reference wavelength ($\lambda_r$) is 17.44$^o$, which matches quite well with the requirement ($\theta_d=17.0194^o$). The achromaticity of deflection is evaluated by the normalized peak shift, defined as $(\theta_{IA}(\lambda)-\theta_{IA}(\lambda_r))/\theta_{IA}(\lambda_r)$. This quantity remains within $\pm$5\% throughout the spectral band ranging from 715nm to 768nm as illustrated in Figure \ref{fig:4_12_summary}(b).   

For the IC state, the observed deflection angles for $\lambda_1$ and $\lambda_2$ are 8.43$^o$ and -4.27$^o$ respectively. These values are in very good agreement with the requirements ($\theta_1$=8.04$^o$ and $\theta_2$=-3.77$^o$). Also, from Figure \ref{fig:4_12_summary}(c), we can see that the reflected light gets deflected towards $\theta_1$ and $\theta_2$ for incident wavelengths in the ranges 745-775nm and 700-735nm, respectively. It means that light within the two ranges will get spatially separated from one another, manifesting wavelength beam splitting. Also, these wavelength ranges align with the operating bands of the other states.

\begin{figure}[!b]
  \centerline{\includegraphics[width=6.5in]{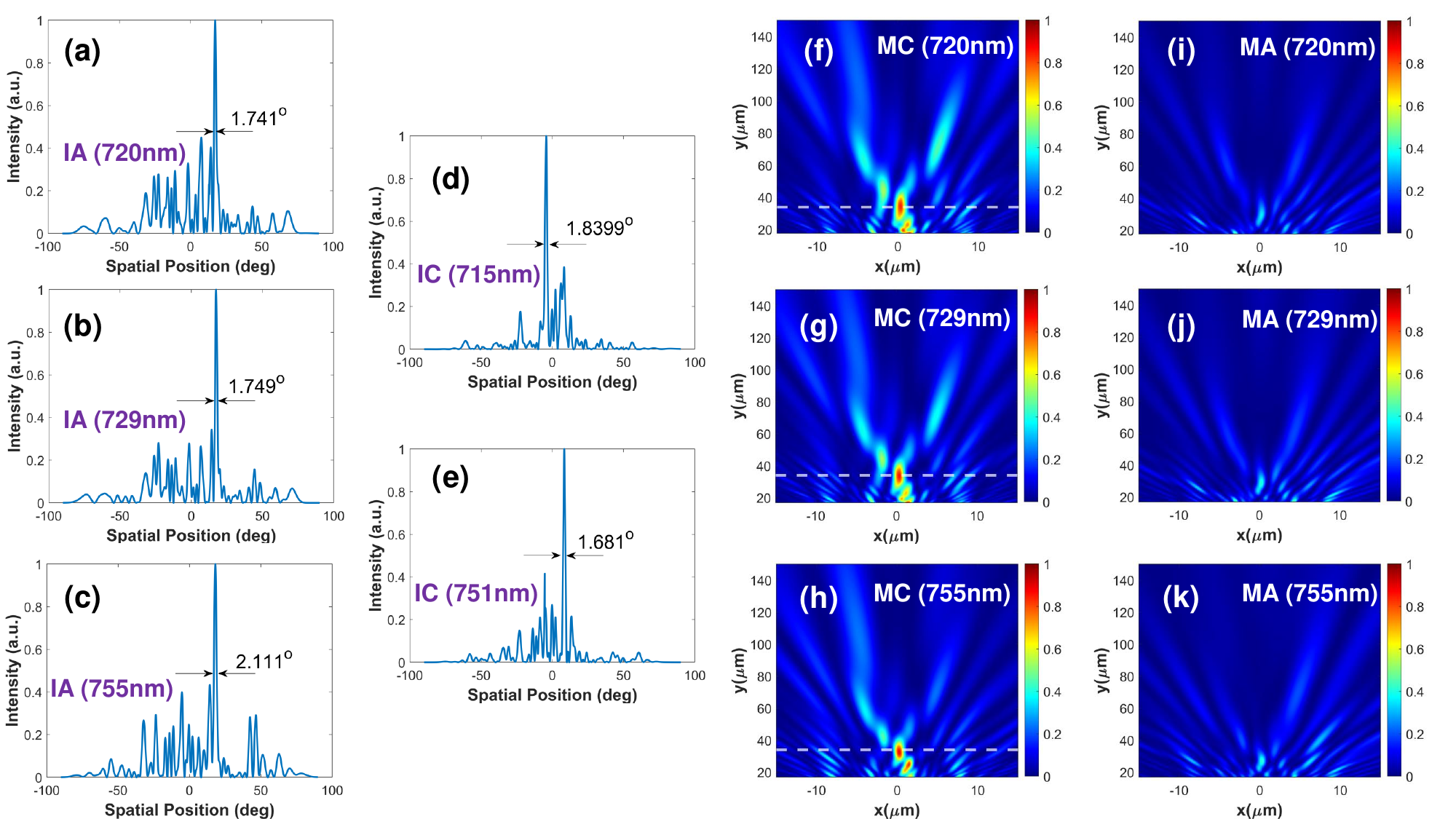}}

\caption{Normalized cross-sectional intensity distributions for the BMFM-1 structure in the \textbf{(a)-(c)} IA, and \textbf{(d)-(e)} IC states at different wavelengths. The arrows mark the corresponding FWHMs. Normalized x-y plane intensity profiles for the \textbf{(f)-(h)} MC, and \textbf{(i)-(k)} MA state at different wavelengths within the operating bandwidth. The intensity values for IA, IC, and MC have been normalized by the respective maxima at that wavelength. For MA in (i)-(k), the color bars were normalized by the maximum intensities of the corresponding MC state to better demonstrate the switching effect. The white dashed lines in (f)-(h) represent the reference wavelength (729.493nm) focal plane positions in the MC state and are used to demonstrate the small focal length shifts in the other two wavelengths. }
\label{fig:4_12_far_field}
\end{figure}

For the MC state, the focal length for the reference wavelength ($\lambda_r$) is 33.964$\mu$m, which is close to the required focal length $f$(=32.6665$\mu$m). The achromaticity of the lensing action is measured by the normalized focal shift, defined as $(f_{MC}(\lambda)-f_{MC}(\lambda_r))/f_{MC}(\lambda_r)$, where $f_{MC}(\lambda)$ is focal length achieved at a wavelength of $\lambda$ in the MC state. This parameter remains within $\pm$5\% over the operating bandwidth between 715nm and 768nm (Figure \ref{fig:4_12_summary}(d)). For the MA state  (or the off state), the broadband absorption is quantified by the ratio of the maximum intensities in the MC and MA states \cite{karim2023reconfigurable}. The MC state is chosen for this definition as it has the lowest reflectance threshold among the on states (IA, IC, and MC). The switching ratio remains above 3dB between 715nm and 775nm, aligning with the operating band of the other states. This threshold of 3dB has been used for differentiating on and off states in previous literature \cite{wang2023multifunctional,ma2022highly}. 

\begin{table}[b!]
\center
\caption{Different optical performance parameters of BMFM-1 and 2.}
\begin{tabular}{p{0.08\linewidth} p{0.42\linewidth} p{0.2\linewidth} p{0.2\linewidth}  }

\hline 
\textbf{State} & \textbf{Performance Matrix} & {\textbf{BMFM-1}} & {\textbf{BMFM-2}}  
\\  
\hline
\textbf{All} & Operating Band & 715-768nm & 700-752nm

\\
\multirow{3}{5em}{\textbf{IA}} & Required deflection angle ($\theta_d$) & 17.0195$^o$ & -5.4511$^o$
\\
&Deflection angle @ $\lambda_r$ & 17.0914$^o$ & -5.368$^o$
\\
& Normalized shift over the operating band & -1.12 to 5\% & 0.07 to 5\%

\\

\\
\multirow{2}{3em}{\textbf{IC}} & Required deflection angles ($\theta_1$, $\theta_2$) & 8.0237$^o$, -3.7697$^o$  & -1.9536$^o$, -7.0567$^o$
\\
& Realized deflection angles & 8.4282$^o$, -4.39$^o$ & -0.8312$^o$, -6.982$^o$

\\
\multirow{3}{5em}{\textbf{MC}} & Required focal length ($f$) & 32.6665$\mu$m & 45.4764$\mu$m
\\
& Focal length @ $\lambda_r$ & 33.964$\mu$m & 39.1491$\mu$m
\\
& Normalized focal shift over the operating band  & 1.65 to -4.54\% &  0.706 to -4.56\%

\\
\hline
\quad
\end{tabular}
\label{table:BMFM}
\end{table}

The aligned operating bandwidth of the BMFM-1 comprises the spectral region from 715nm to 768nm. This band was chosen by ensuring a normalized deflection angle and focal length shift within $\pm$5\% for the IA and MC states, respectively, and an intensity switching ratio between MC and MA of not less than 3dB. Also, the two wavelengths (715nm and 751.623nm) for the beam-splitting operation in IA state are also within this band. The far-field distributions for different states of BMFM-1 at different wavelengths are also shown in Figure \ref{fig:4_12_far_field}. The achromatic deflection and focusing in the IA (Figure \ref{fig:4_12_far_field}(a)-(c)) and MC (Figure \ref{fig:4_12_far_field}(f)-(h)) state and the splitting of the reflected light at 715nm and 751.623nm towards different directions in the IC state (Figure \ref{fig:4_12_far_field}(d)-(e)) are visible in the farfield patterns. The disappearance of the reflected beam intensity for the MA state in Figure \ref{fig:4_12_far_field}(i)-(k) represents the broadband absorption in this state. 

We have also evaluated the optical response of another structure (BMFM-2). The meta-device was inverse designed using the TNN architecture for the parameters: $\theta_d=-5.4512^o, \theta_1=-1.9536^o, \theta_2=-7.0567^o$, and $f=45.4764\mu m$. The simulated responses of the optimized meta-units are compared with the required values in Figure S9 of the SI. Both sets of values are in very good agreement. The detailed analysis of the optical response for the four states of the BMFM-2 is shown in Figure S7 and S8 of the SI. The results show that the structure has an aligned operating bandwidth of 52nm (from 700nm to 752nm). The required and realized values of different performance parameters for BMFM-1 and 2 in the four states are summarized in Table \ref{table:BMFM}. Both the inverse-designed BMFMs closely fulfill the requirements over the entire bandwidth. The design parameters for the two structures were chosen at random from different probable ranges to demonstrate the diversity of the inverse design methodology. Such broadband metasurface with four or more functionalities has not been reported before in any spectral band. Table \ref{table:review} compares the BMFMs designed in this work with previously reported multifunctional metasurfaces. Previous works have mainly been restricted to the THz and GHz operating frequencies, and also are not broadband in nature, further iterating the significance of this work.

\begin{table}[h!]
\center
\caption{Comparison of our BMFM structures with previously reported multifunctional meta-devices.}
\begin{tabular}{p{0.25\linewidth} p{0.25\linewidth} p{0.25\linewidth} p{0.11\linewidth}}

\hline 
\textbf{Switching mechanism} & \textbf{Operating wavelength range} & \textbf{Functionalities} &  \textbf{Reference} 
\\
 
\hline

PIN diode & 4.62 to 13.56 GHz (Two of the functionalities are only realized at 10GHz) & Different combinations of reflection and transmission of x and y polarized light & \cite{li2019design}

\\
\\
Polarization, frequency and incidence direction switched & 8,10,17 GHz (different functionality at different wavelengths) & beam focusing lens, holography, beam steering, diverging lens & \cite{zhu2021four}

\\
\\
Individually addressable $\mathrm{VO_2}$ based meta-atoms & 1.69THz &  Absorption, vortex beam generation, lensing, beam deflection & \cite{yang2022switchable}

\\
\\
$\mathrm{VO_2}$ and Graphene & Different functions at different THz wavelengths & circular dichroism, x to y and linear to circular conversion, absorption, asymmetric transmission & \cite{xu2022multifunctional}

\\
\\
Multiplexed metasurfaces & Different functions at different GHz wavelengths & perfect transmission, reflection, absorption, transmission phase modulation & \cite{ma2022highly}

\\
\\
Phase transition of $\mathrm{VO_2}$ and $\mathrm{Sb_2S_3}$ & 715-768nm (BMFM-1), 700-752nm (BMFM-2) & Achromatic beam deflection, wavelength beam splitting, achromatic focusing, broadband absorption & \textbf{This work}
\\
\hline
\quad
\end{tabular}
\label{table:review}
\end{table}

The findings of this work show the vast potential of artificial intelligence to produce active multi-state metasurfaces, leading the way for miniaturized complex nanophotonics systems in different design bands. Also, the excellent performance of our proposed metasurfaces through TNN inverse design means they can find direct applications in medical imaging technologies like Optical Coherence Tomography (OCT), used for non-traditional biopsy of human tissue. The currently available OCT systems are not portable owing to the bulky base unit \cite{ong2022advances}. The base unit performs beam splitting, focusing, beam deflection, and steering along with other functionalities \cite{aumann2019optical}. Also, recently, there has been a growing interest in the visible wavelength for OCT systems to achieve higher axial resolution. Such vis-OCT systems will greatly benefit from our BMFMs due to their ability to integrate several relevant functionalities in the same meta-structure at visible frequencies. This will pave the way for portable vis-OCT systems for on-site non-invasive imaging and oximetry of the human retina \cite{song2020visible}.  

\section{Conclusion}
In this work, we report broadband multifunctional metasurfaces capable of demonstrating four distinct functionalities - achromatic deflection, wavelength beam splitting, achromatic focusing, and broadband absorption in the visible range. The meta-atoms combine two PCMs, $\mathrm{VO_2}$ and $\mathrm{Sb_2S_3}$, in a multistage configuration to produce four distinct broadband states utilizing the cavity length modulation effect from PCM state transition, overcoming the insignificant optical contrast of PCM states in the visible range. This distinct feature allows us to incorporate both switchable amplitude and phase based functionalities in the same structure. We used a TNN architecture capable of turning the complex inverse design process, involving several frequency and wavelength-dependent requirements in each state, into a regression problem. The two forward networks used for predicting the reflection phase and amplitude of an input meta-atom within the predefined design space were connected in tandem with the inverse network during training, avoiding the non-convergence problem. The excellent performance of all three networks on the test data set proves the efficiency of the supervised learning using a moderately sized data set. Two metasurfaces were inversely designed for different spectral requirements in the four states. Numerical simulations of the two test cases demonstrate broadband operation in all four states with overlapping operating bands, a feature not realized in previous literature, while closely matching the design specifications of the four distinct functions. Our findings in this work show the immense potential of active metasurfaces in producing complex broadband multifunctionality by smart integration of artificial intelligence and PCMs and are expected to open new avenues for active multifunctional devices for next-generation nano-photonic platforms.

\vspace{5mm}
\large{\textbf{Acknowledgement}}\\
\small{The authours gratefully acknowledge the finanical support of Bangladesh University of Engineering and Technology through the RISE Internal Research Grant
(Application ID 2023-02-006).}

\vspace{5mm}
\large{\textbf{Supporting Information}}\\
\small{Supporting Information Available: Methods, Optical Properties of a single-stage unit cell, Additional details of the neural networks, Further analysis of the BMFMs (PDF).}

\vspace{5mm}
\large{\textbf{Data Availability Statement}}\\
\small{Data underlying the findings of this work are not public at this time but will be made available upon reasonable request to the authors.}

\bibliographystyle{elsarticle-num-names} 
\bibliography{Main2}





\end{document}


\begin{frontmatter}



\title{Supporting Information for: Synergizing Deep Learning and Phase Change Materials for Four-state Broadband Multifunctional Metasurfaces in the Visible Range} 


\author[label1,label3,label4]{Md. Ehsanul Karim} 
\author[label2,label4]{Md. Redwanul Karim} 
\author{Sajid Muhaimin Choudhury\fnref{label1}}
\ead{sajid@eee.buet.ac.bd}

\affiliation[label1]{organization={Department of Electrical and Electronic Engineering, Bangladesh University of Engineering and Technology},
            city={Dhaka},
            postcode={1205}, 
            country={Bangladesh}}

\affiliation[label2]{organization={Department of Computer Science and Engineering, Bangladesh University of Engineering and
Technology},
            city={Dhaka},
            postcode={1205}, 
            country={Bangladesh}}

\affiliation[label3]{organization={Department of Electrical and Electronic Engineering, BRAC University},
            city={Dhaka},
            postcode={1212}, 
            country={Bangladesh}}

\affiliation[label4]{Co-first authors with equal contributions}




\end{frontmatter}


\section{Methods}
\subsection{Numerical Modelling}
The optical characterization of the unit cells and the final metasurfaces has been conducted in the commercial software Lumerical FDTD Solutions. For the unit cell simulations, an x-polarized plane wave source propagating along the negative y direction was incident on the meta-unit, illustrated by the directions of electric field ($E$) and propagation (\textit{k}) vectors, respectively in Figure 1(b) of the main text. The reflectance was calculated using a 2d frequency domain power monitor in the simulation environment. Periodic boundary conditions have been used for the x direction, and the y direction contains Perfectly Matched Layer (PML) boundaries. For the broadband multifunctional metasurface (BMFM) structure an x-polarized electromagnetic plane wave propagating in the -y direction is incident, as shown in Figure 6(a) of the main text. PML boundary conditions were used in both the x and y directions \cite{balli2021ultrabroadband}. 

\begin{figure}[!h]
  \centerline{\includegraphics[width=6.5in]{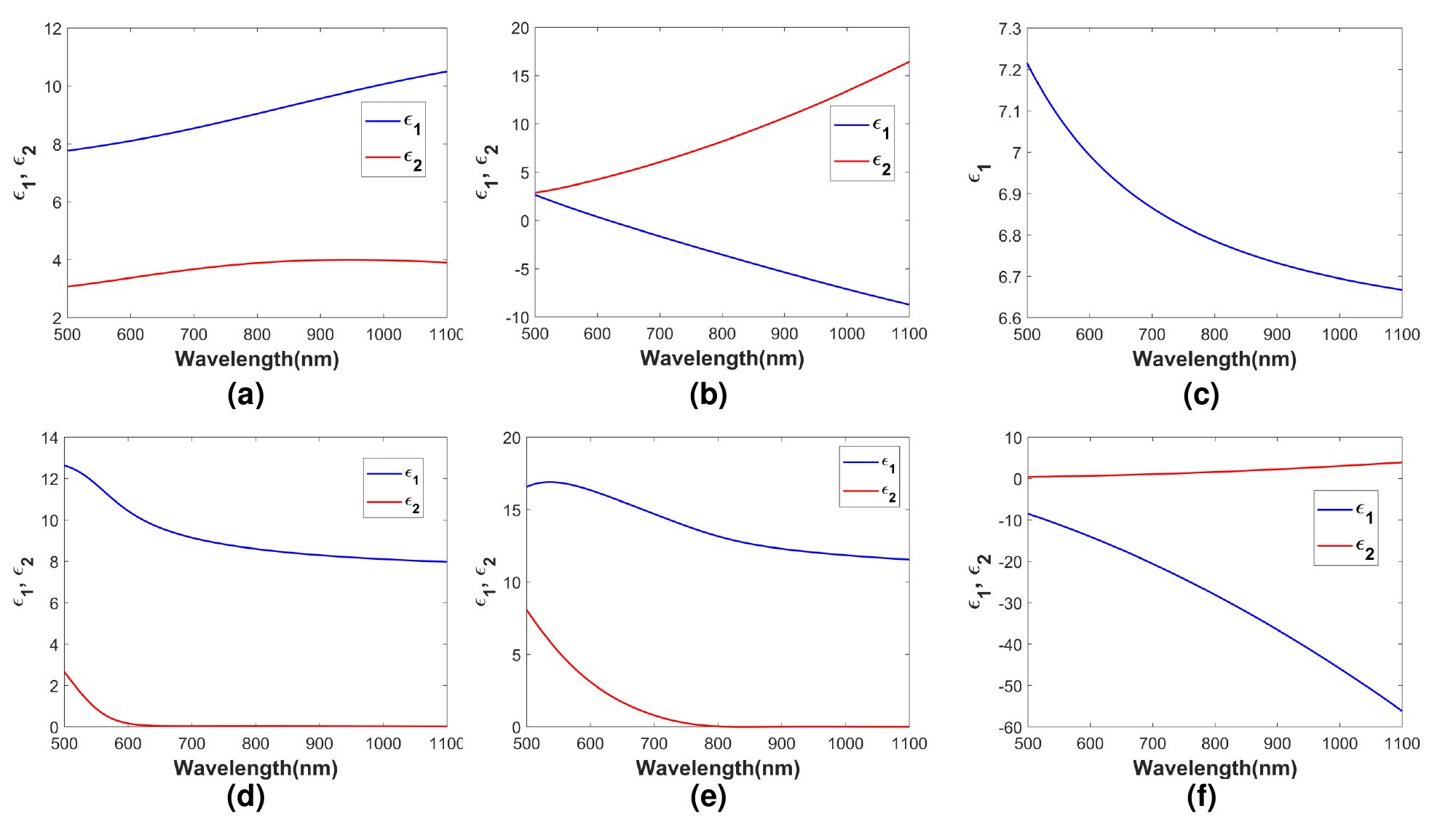}}

\caption{The real ($\epsilon_1$) and imaginary ($\epsilon_2$) parts of relative permittivity for \textbf{(a)} insulating and \textbf{(b)} metallic $\mathrm{VO_2}$ \cite{dicken2009frequency}, \textbf{(c)} 6H-SiC \cite{wang20134h}, \textbf{(d)} amorphous and \textbf{(e)} crystalline $\mathrm{Sb_2S_3}$ \cite{delaney2020new}, and \textbf{(f)} Ag \cite{palik1998handbook} used for the numerical simulations. The imaginary component of permittivity for 6H-SiC is small in our operating band \cite{xie2003first} and hence has been considered zero.}
\label{fig:material}
\end{figure}

The optical properties of $\mathrm{VO_2}$ \cite{dicken2009frequency}, $\mathrm{Sb_2S_3}$ \cite{delaney2020new}, 6H-SiC \cite{wang20134h,xie2003first}, and Silver \cite{palik1998handbook} have been taken from previously reported values (Figure \ref{fig:material}). The device will operate at room temperature and at above 68$\mathrm{^o}$C for the insulating and metallic state of $\mathrm{VO_2}$, respectively. However, the thermo-optic coefficients of the other materials (Ag, SiC, $\mathrm{Sb_2S_3}$) are negligible within this operating temperature range \cite{xu2014temperature,sundari2013temperature,fang2021non} range and hence have been neglected in our simulations.

\subsection{Calculation of loss components for IDN}
The loss component for reflectivity ($T_i'$) is defined as the error of the predicted reflectivity (from RPN) not being able to satisfy the threshold in different states within the intended operating bandwidth. This is calculated from the following set of equations

\begin{equation}
\label{eqn:RL_IA}
    T_{i,IA}' = \begin{cases} 
          0.4-R_{i,IA}' & R_{i,IA}'<0.4 \\
           0 & \mathrm{otherwise} \\
       \end{cases}
\end{equation}

\begin{equation}
\label{eqn:RL_IC}
    T_{i,IC}' = \begin{cases} 
          0.4-R_{i,IC}' & R_{i,IC}'<0.4 \\
           0 & \mathrm{otherwise} \\
       \end{cases}
\end{equation}

\begin{equation}
\label{eqn:RL_MC}
    T_{i,MC}' = \begin{cases} 
          0.2-R_{i,MC}' & R_{i,MC}'<0.2 \\
           0 & \mathrm{otherwise} \\
       \end{cases}
\end{equation}

\begin{equation}
\label{eqn:RL_MA}
    T_{i,MA}' = \begin{cases} 
          R_{i,MA}'-0.1 & R_{i,MA}'>0.1 \\
           0 & \mathrm{otherwise} \\
       \end{cases}
\end{equation}

\begin{equation}
\label{eqn:MAE_PPN}
    T_{i}' = [T_{i,IA}',T_{i,IC}',T_{i,MC}',T_{i,MA}']
\end{equation}
Where $R_{i,j}'$ is a 13x1 vector of the RPN predicted reflectance spectra between 698.65 and 763.19nm for the \textit{j}th state ($j=IC,IA,MC,MA$) of the \textit{i}th meta-atom in data set.

The phase requirement function vector ($F$) is constructed based on the specific phase gradient meta-optics equation applicable to the individual states. For achromatic deflection of the reflected beam in the IA state to a desired direction, $\theta_d$, the relative phase ($\Phi_{IA}(r,\omega)$) required at any point $r$ of the metasurface is given as \cite{forouzmand2016tunable}

\begin{equation}
\label{eqn:phase_IA}
    \Phi_{IA}(r,\omega)=\frac{\omega}{c}rsin(\theta_d)
\end{equation}
where $\omega$ and $c$ are the angular frequency and speed of the incident light, respectively. This equation can be expanded using the well-known Taylor series expansion about a reference frequency $\omega_r$
\begin{equation}
\label{eqn:phase_expansion_IA}
    \Phi_{IA}(r,\omega)=\Phi_{IA}(r,\omega_r)+\frac{\partial \Phi_{IA}(r,\omega)}{\partial \omega}\bigg|_{\omega=\omega_r}(\omega-\omega_r)+\frac{\partial^2 \Phi_{IA}(r,\omega)}{2\partial^2 \omega}\bigg|_{\omega=\omega_r}(\omega-\omega_r)^2+...
\end{equation}
All the terms can be calculated using Eq. \ref{eqn:phase_IA}. Theoretically, all these terms need to be matched at every point of the metasurface over the entire wavelength range to achieve achromatic deflection. However, only considering the first three terms has yielded wavelength-independent operation in previous works \cite{jia2021achromatic}. The first and second order derivatives around $\omega_r$ in Eq. \ref{eqn:phase_expansion_IA} are called group delay ($gd_{IA}(r)$) and group delay dispersion ($gdd_{IA}(r)$) respectively \cite{chen2019broadband}.

The inverse designed metasurface is expected to produce wavelength beam splitting in the IC state, that is, reflecting incident light of two different wavelengths ($\lambda_1$, $\lambda_2$) into two different directions ($\theta_1$ and $\theta_2$). For this, the reflection phase at the two target frequencies should satisfy the following mathematical relations
\begin{equation}
\label{eqn:phase_IC1}
    \Phi_{IC}(r,\omega_1)=\frac{\omega_1}{c}rsin(\theta_1)
\end{equation}

\begin{equation}
\label{eqn:phase_IC2}
    \Phi_{IC}(r,\omega_2)=\frac{\omega_2}{c}rsin(\theta_2)
\end{equation}

The required relative phase ($\Phi_{MC}(r,\omega)$) for the MC state at any point ($r$) on the metasurface with respect to the center for a focal length of $f$ is given by \cite{cheng2019broadband}
\begin{equation}
\label{eqn:phase}
    \Phi_{MC}(r,\omega)=-\frac{\omega}{c}(\sqrt{r^2+f^2}-f)
\end{equation}
For achromatic focusing, Eq. \ref{eqn:phase} needs to be satisfied at every point on the metasurface through the design bandwidth in the MC state. Taylor series theorem can be used to expand Eq. \ref{eqn:phase} about a reference frequency of $\omega_r$ 
\begin{equation}
\label{eqn:phase_expansion}
    \Phi_{MC}(r,\omega)=\Phi_{MC}(r,\omega_r)+\frac{\partial \Phi_{MC}(r,\omega)}{\partial \omega}\bigg|_{\omega=\omega_r}(\omega-\omega_r)+\frac{\partial^2 \Phi_{MC}(r,\omega)}{2\partial^2 \omega}\bigg|_{\omega=\omega_r}(\omega-\omega_r)^2+...
\end{equation}
For this state as well, we ignore the higher order terms beyond the second derivative \cite{chen2018broadband}. The first and second order derivative terms at $\omega_r$ are called group delay ($gd_{MC}(r)$) and group delay dispersion ($gdd_{MC}(r)$) respectively and can be calculated from Eq. \ref{eqn:phase}. Since we have chosen the final design bandwidth to ensure a linear phase profile in all the states, the delay dispersion terms for both IA and MC states can be ignored \cite{qiu2024neural}.

Based on the discussion above, the meta-atoms of the inverse designed meta-device must satisfy the reflection phase and group delay in the IA and MC states around a reference wavelength ($\lambda_r=2\pi c/\omega_r$) over the design bandwidth at all the points of the metasurface. For our design, we have chosen $\lambda_r$=729.493nm and the bandwidth from 698.65 to 763.19nm. For the IC state, the same meta-atoms should satisfy the spatial phase distributions for the two target wavelengths ($\lambda_1$=708.629 and $\lambda_2$=751.623nm), given by Eq. \ref{eqn:phase_IC1} and \ref{eqn:phase_IC2}. So, the phase requirement function ($F$) of the \textit{i}th structure of the data set is constructed as $F_i=[\phi_{i,IA}(\lambda_r), \phi_{i,IC}(\lambda_1), \phi_{i,IC}(\lambda_2), \phi_{i,MC}(\lambda_r), GD_{i,IA}, GD_{i,MC}]$. Also, it should be noted that we have used different notations for required ($\Phi,gd$) and ground truth ($\phi, GD$) phase and group delays in this discussion for clarity.   

\newpage
\section{Optical Properties of a single-stage unit cell}
\label{sec:unit cell}

\begin{figure}[!h]
  \centerline{\includegraphics[width=6.5in]{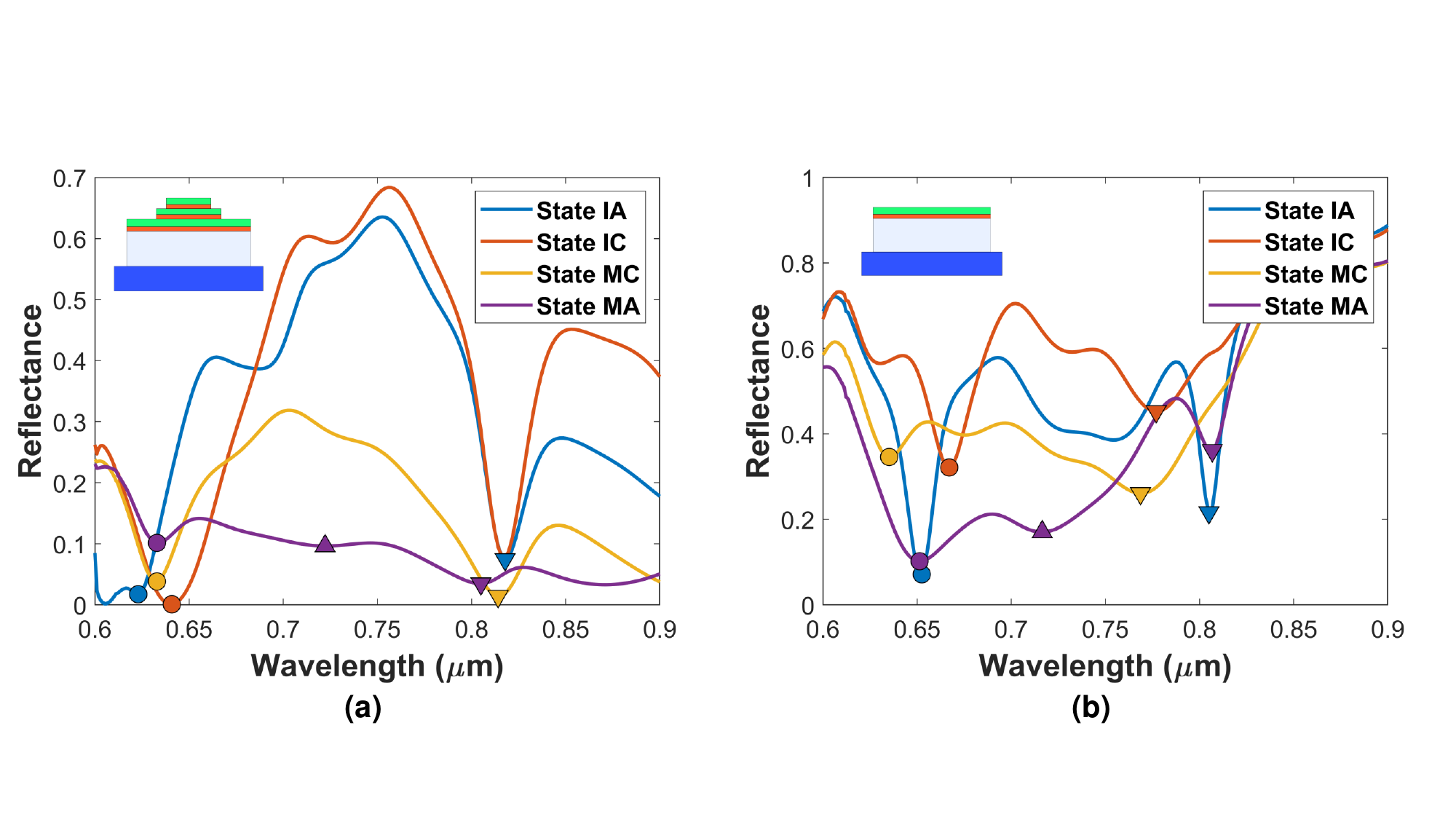}}

\caption{The reflection spectra for four different states of the representative unit cell of the main text with \textbf{(a)} three stages and \textbf{(b)} a single stage. The insets show the x-y plane cross-section of the corresponding unit cell structures. Both the structures have the same geometrical parameters as in Figure 2 of the main text. The symbols $\circ$, $\bigtriangledown$, and $\bigtriangleup$ indicate the Mode 1, 2, and 3 wavelength positions, respectively, at the respective states. The broadband amplitude switching capability between MA and MC vanishes when the structure has a single stage.}
\label{fig:unit_cell}
\end{figure}

\newpage
\section{Additional details of the neural networks}

\begin{table}[h!]
\center
\caption{Settings of some hyper-parameters of the three neural networks}
\begin{tabular}{p{0.25\linewidth} p{0.25\linewidth} p{0.25\linewidth} p{0.25\linewidth}}

\hline 
\textbf{Hyperparameters} & \textbf{PPN} & \textbf{RPN} & \textbf{IDN}
\\
 
\hline
Initialization & Xavier Initialization\cite{xavierInitialization} & Xavier Initialization & Xavier Initialization 

\\
Hidden layers & 8 & 7 & 9

\\
Nodes &  11, 2400, 2000, 1600, 1200, 1000, 800, 600, 400, 45 & 11, 2000, 1600, 1200, 1000, 800, 600, 400, 160 & 10, 2000, 1600, 1200, 1000, 800, 600, 400, 200, 100, 11

\\

Loss function & MAE & MAE & Weighted MAE

\\
Optimizer & Adam & Adam & Adam

\\

Batch size & 20 & 20 & 32

\\
Activation function & Leaky Relu(0.1) & Leaky Relu(0.1) & Leaky Relu(0.1)

\\
Initial Learning rate & 0.0005 & 0.0001 & 0.0001

\\

\hline
\quad
\end{tabular}
\label{table:review}
\end{table}

\begin{figure}[!h]
  \centerline{\includegraphics[width=6.5in]{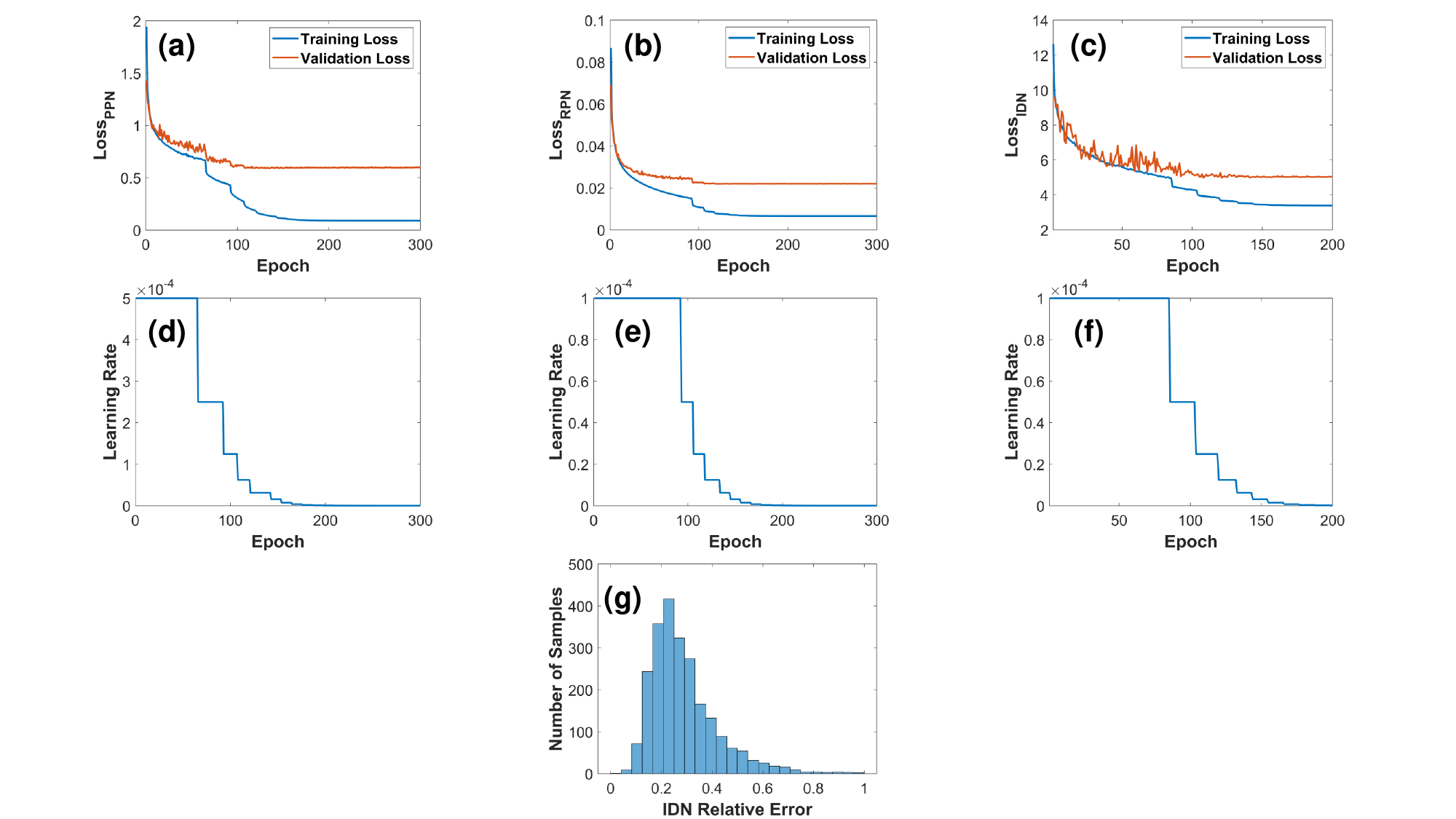}}

\caption{The learning and validation loss curves for \textbf{(a)} PPN, \textbf{(b)} RPN, and \textbf{(c)} IDN. Dynamic learning rate for \textbf{(d)} PPN, \textbf{(e)} RPN, and \textbf{(f)} IDN. \textbf{(g)} Relative error distributions of the IDN for the 2328 test samples. }
\end{figure}

\begin{figure}[!h]
  \centerline{\includegraphics[width=6.5in]{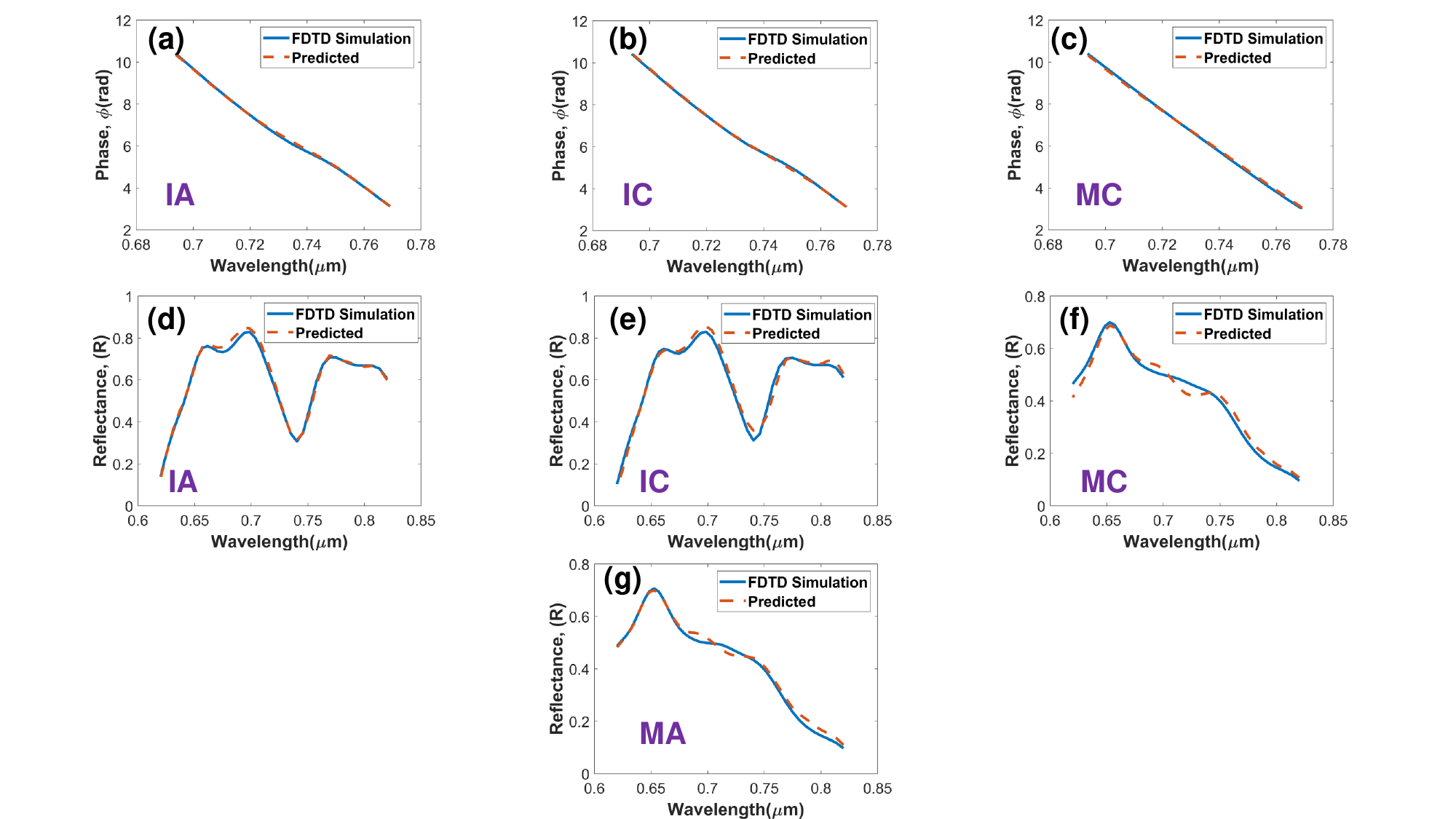}}

\caption{\textbf{(a)-(c)} Reflection phase ($\phi$) and \textbf{(d)-(g)} Reflectance ($R$) spectra of different states of a randomly selected unit cell from the test data set, calculated from FDTD simulations and predicted from PPN and RPN respectively. The structural parameters input to the neural networks for this test case are $w_1$=516.9677nm, $h$=70.32359nm, $d$=228.0329nm, $t_1$=252.03937nm, $v_1$=22.80398nm, $v_2$=21.69389nm, $v_3$=14.563346nm, $x_1$=0.18387112, $x_2$=0.13597298, $y_1$=0.9895365, and $y_3$=1.0169164. }
\label{fig:forward}
\end{figure}

\begin{figure}[!h]
  \centerline{\includegraphics[width=6.5in]{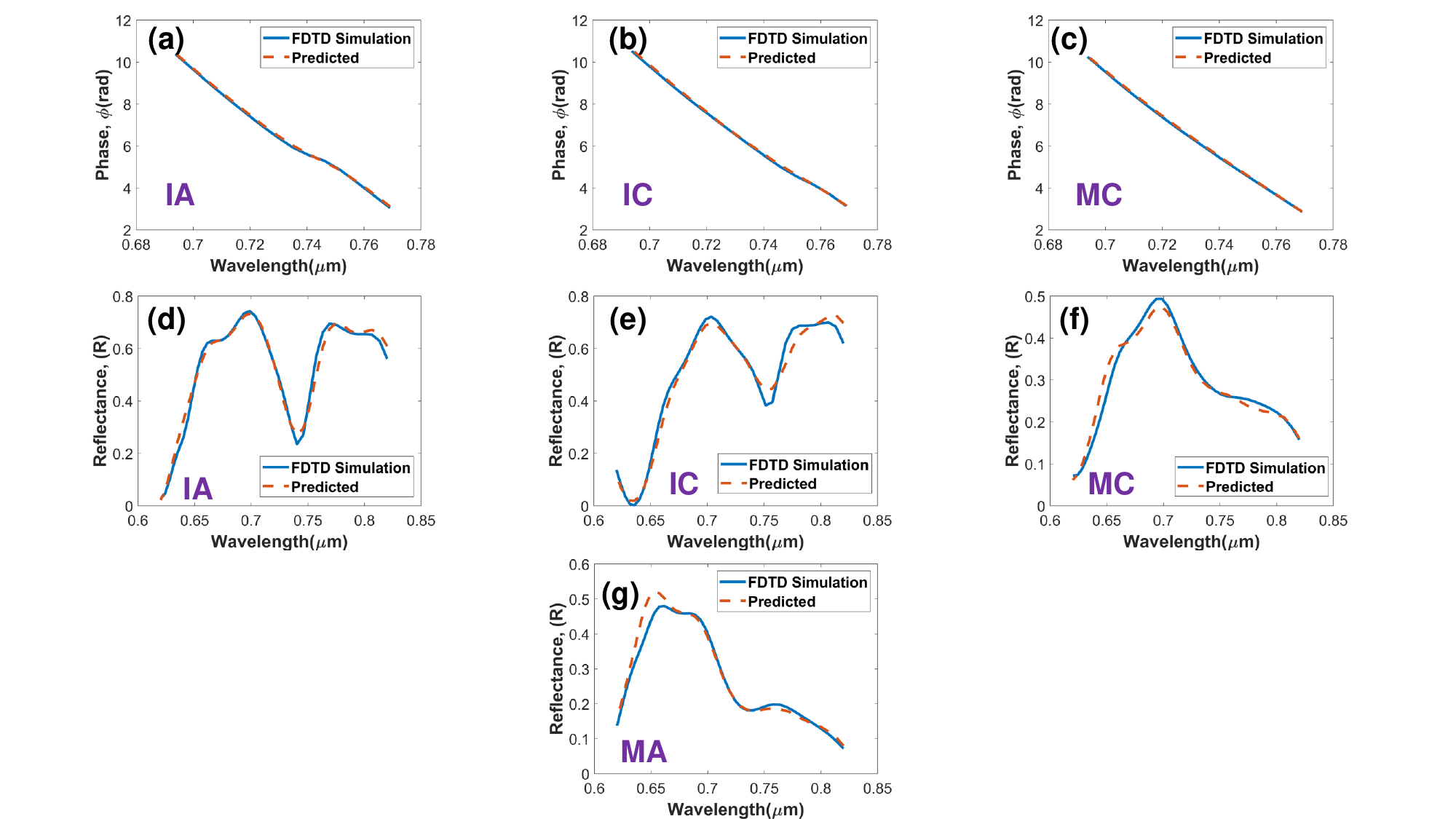}}

\caption{\textbf{(a)-(c)} Reflection phase ($\phi$) and \textbf{(d)-(g)} Reflectance ($R$) spectra of different states of a randomly selected unit cell from the test data set, calculated from FDTD simulations and predicted from PPN and RPN respectively. The structural parameters input to the neural networks for this test case are $w_1$=516.24695nm, $h$=92.14034nm, $d$=207.81909nm, $t_1$=238.82242nm, $v_1$=14.516156nm, $v_2$=22.883387nm, $v_3$=13.8248415nm, $x_1$=0.47292465, $x_2$=0.09616624, $y_1$=0.7904597, and $y_3$=1.3373394. }
\label{fig:forward2}
\end{figure}

\newpage
\FloatBarrier
\section{Further analysis of the BMFMs}
\begin{figure}[!h]
  \centerline{\includegraphics[width=6.5in]{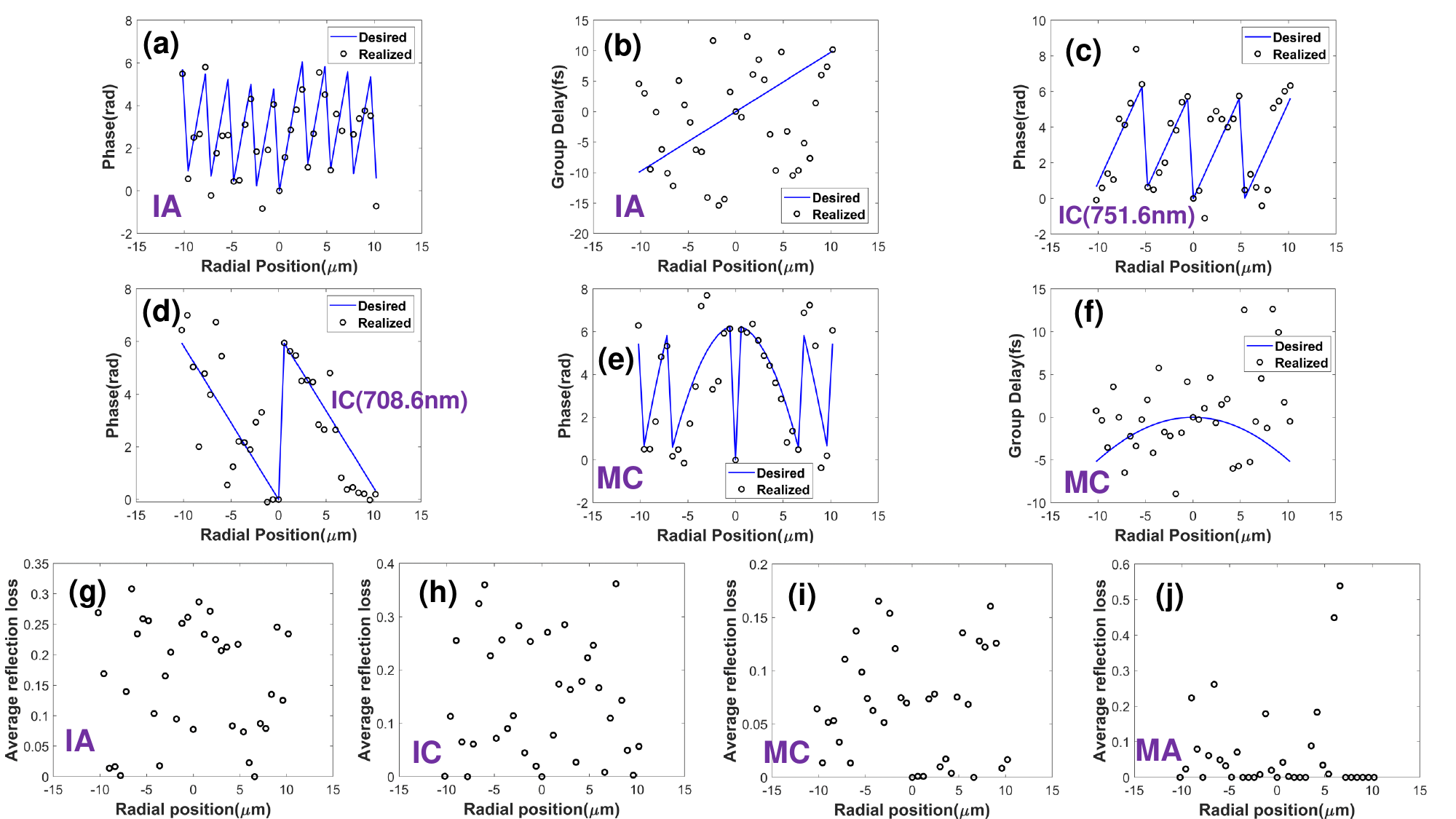}}

\caption{\textbf{(a)-(f)}The desired and realized values of relative phase and group delays for the IA, IC, and MC states of the BMFM-1 structure as a function of position on the meta-device. The required values were calculated using Eq. \ref{eqn:phase_IA}-\ref{eqn:phase_expansion}, respectively, for design requirements of $\theta_d=17.0194^o, \theta_1=8.0237^o, \theta_2=-3.7697^o, f=32.6665\mu m, \lambda_1=751.623nm, \lambda_2=708.629nm.$ They construct the 35 phase function vectors (\textit{F}) (corresponding to the 35 positions on the metasurface) input to the IDN for generating the meta-atoms. The realized values are the FDTD simulation responses of these inverse-designed unit cells. \textbf{(g)-(h)} average reflection losses of the unit cells as a function of their radial position on the meta-device. They are calculated from Eq. \ref{eqn:RL_IA}-\ref{eqn:RL_MA}  using the reflectance spectra of the unit cells from FDTD simulations.}
\label{fig:4_12_requirements}
\end{figure}

\begin{figure}[!h]
  \centerline{\includegraphics[width=6.5in]{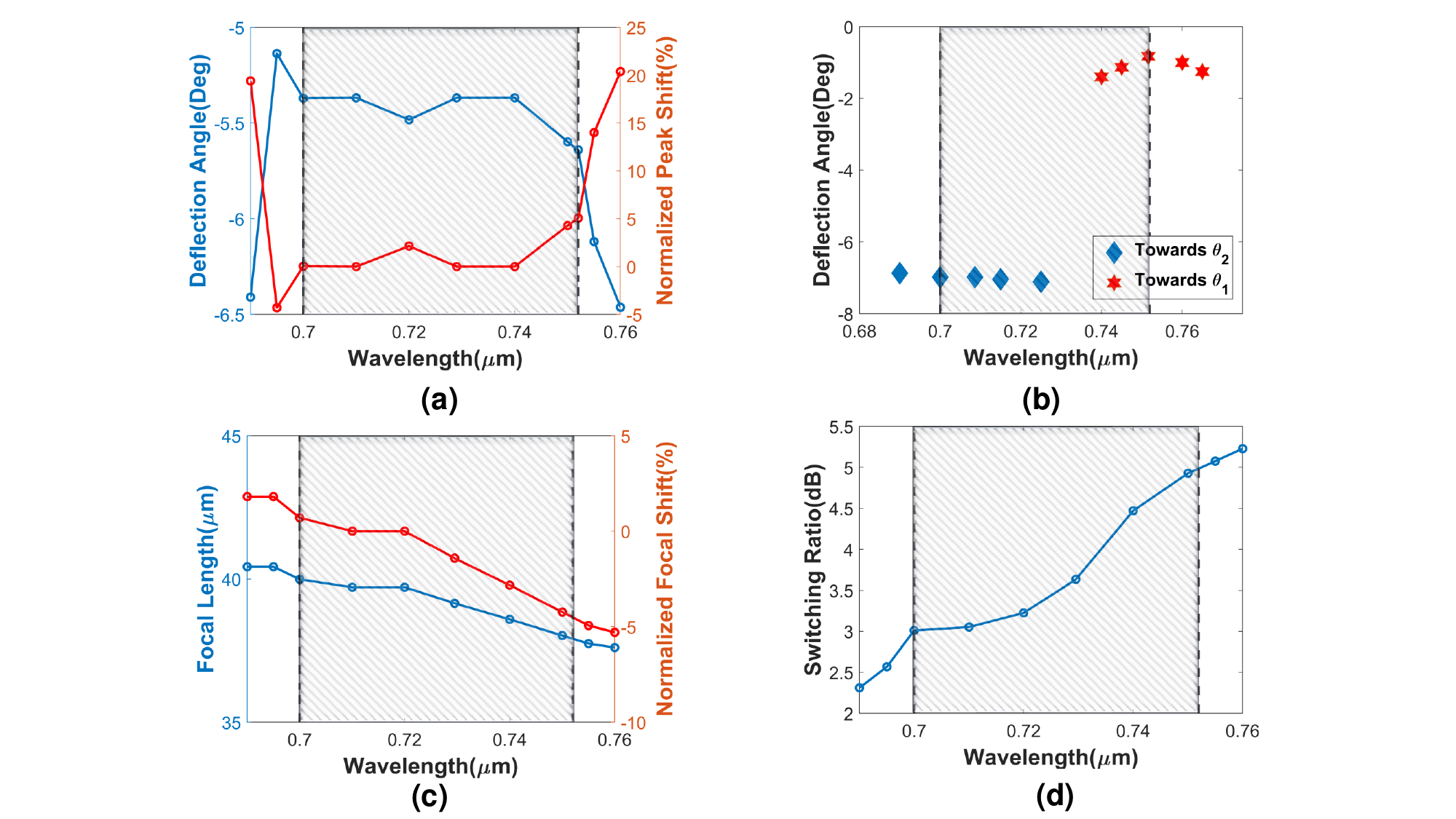}}

\caption{Different optical parameters for the four states of BMFM-2, inverse designed for $\theta_d=-5.4512^o, \theta_1=-1.9536^o, \theta_2=-7.0567^o, f=45.4764\mu m, \lambda_1=751.623nm, \lambda_2=708.629nm.$ \textbf{(a)} Variation of deflection angle and relative angular position shift normalized to the reference wavelength of 729.493nm with incident light frequency in the IA state. \textbf{(b)} Spatial splitting positions of reflected light at two wavelengths in the IC state. Variation of \textbf{(c)} Focal length and normalized focal shift with respect to that in 729.493nm in the MC state, \textbf{(d)} Intensity switching ratio between MC and MA state as a function of wavelength. The hatched regions represent the operating wavelength between 700nm (=$\lambda_L$) and 752nm (=$\lambda_H$). }
\label{fig:4_5_summary}
\end{figure}

\begin{figure}[!h]
  \centerline{\includegraphics[width=6.5in]{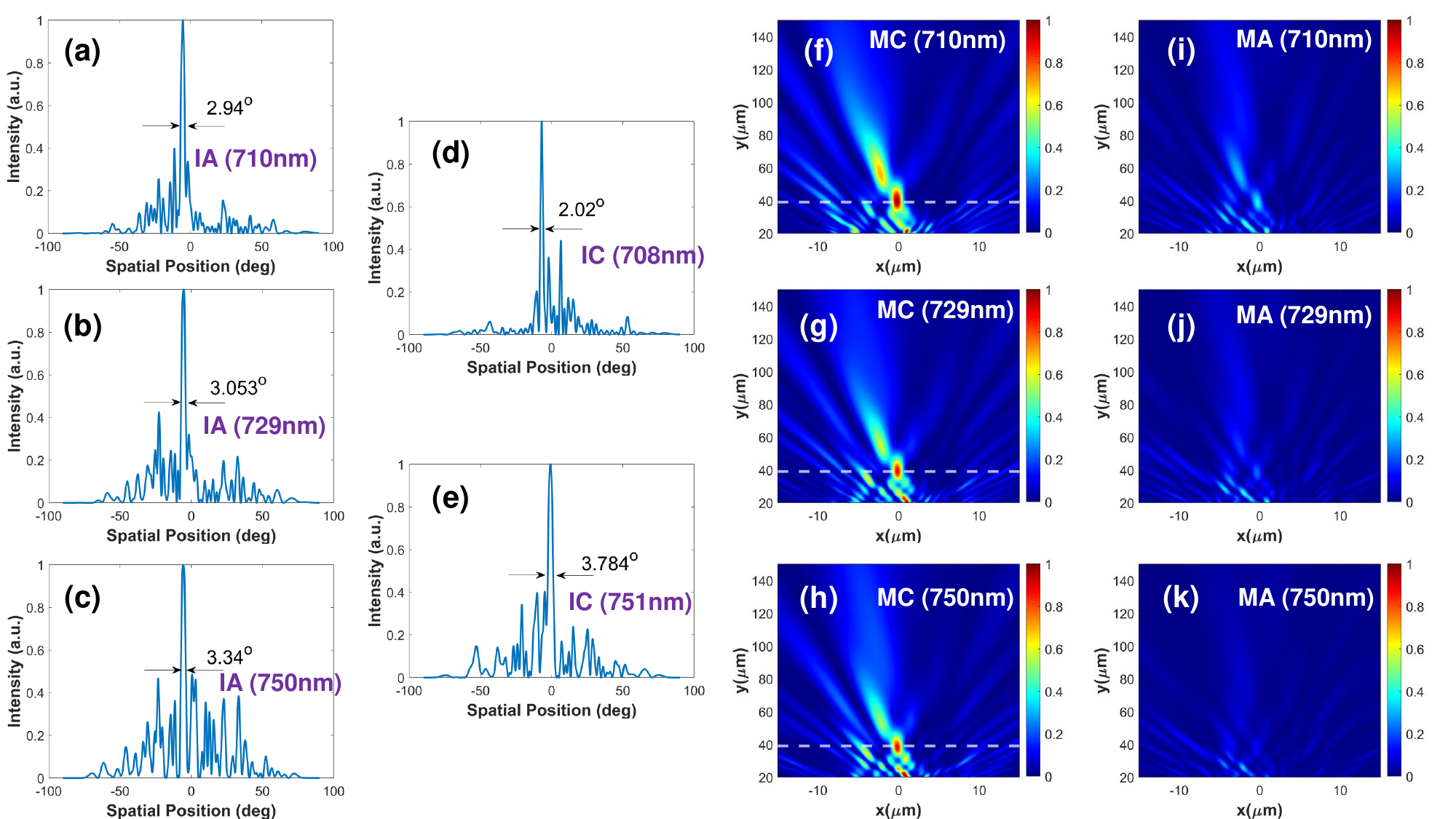}}

\caption{Normalized cross-sectional intensity distributions for the BMFM-2 structure in the \textbf{(a)-(c)} IA, and \textbf{(d)-(e)} IC states at different wavelengths. The arrows mark the corresponding FWHMs. Normalized x-y plane intensity profiles for the \textbf{(f)-(h)} MC, and \textbf{(i)-(k)} MA state at different wavelengths within the operating bandwidth. The intensity values for IA, IC, and MC have been normalized by the respective maxima at that wavelength. For MA in (i)-(k), the color bars were normalized by the maximum intensities of the corresponding MC state to better demonstrate the switching effect. The white dashed lines in (f)-(h) represent the reference wavelength (729.493nm) focal plane positions in the MC state and are used to demonstrate the small focal length shifts in the other two wavelengths. }
\label{fig:4_5_far_field}
\end{figure}

\begin{figure}[!h]
  \centerline{\includegraphics[width=6.5in]{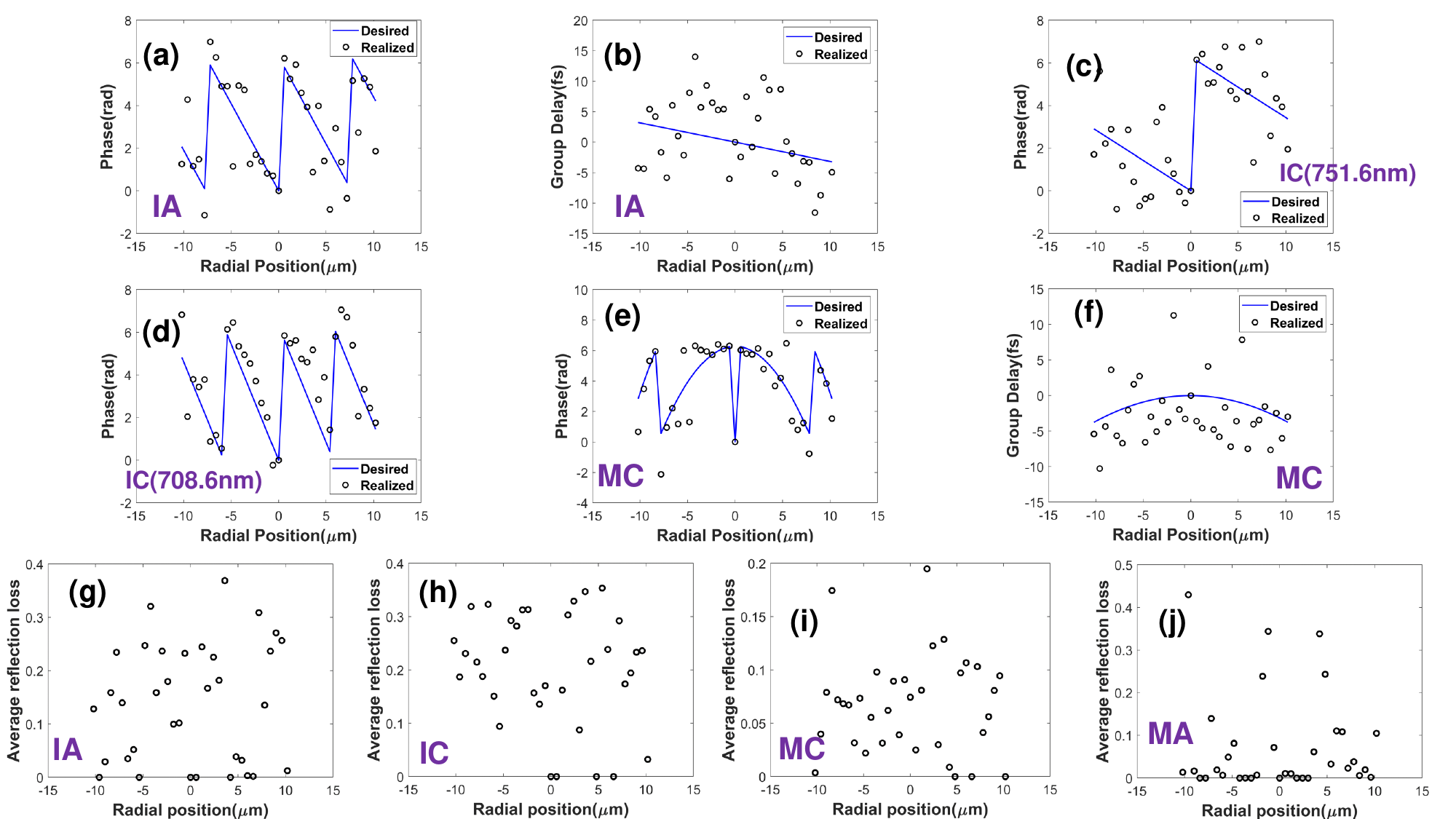}}

\caption{\textbf{(a)-(f)} The desired and realized values of relative phase and group delays for the IA, IC, and MC states of the BMFM-2 structure as a function of position on the meta-device. The required values were calculated using Eq. \ref{eqn:phase_IA}-\ref{eqn:phase_expansion}, respectively, for design requirements of $\theta_d=-5.4512^o, \theta_1=-1.9536^o, \theta_2=-7.0567^o, f=45.4764\mu m, \lambda_1=751.623nm, \lambda_2=708.629nm.$ They construct the 35 phase function vectors (\textit{F}) (corresponding to the 35 positions on the metasurface) input to the IDN for generating the meta-atoms. The realized values are the FDTD simulation responses of these inverse-designed unit cells. \textbf{(g)-(h)} average reflection losses of the unit cells as a function of their radial position on the meta-device. They are calculated from Eq. \ref{eqn:RL_IA}-\ref{eqn:RL_MA}  using the reflectance spectra of the unit cells from FDTD simulations. }
\label{fig:4_5_requirements}
\end{figure}

\newpage
\FloatBarrier


